\documentclass[twocolumn,tighten]{aastex631}
\usepackage{tabto}
\usepackage{lineno}
\usepackage{capt-of}

\shorttitle{VLA Observations of CIZA J0107.7+5408}
\shortauthors{Schwartzman et al.}

\graphicspath{{./}{Figures/}}

\begin{document}

\title{Multi-frequency Radio Observations of the Dissociative Cluster Merger CIZA J0107.7+5408}

\author[0000-0002-6454-861X]{Emma Schwartzman}
\affiliation{U.S. Naval Research Laboratory, 4555 Overlook Ave SW, Washington, DC 20375, USA}
\affiliation{Department of Physics and Astronomy, George Mason University, 4400 University Drive, MSN 3F3, Fairfax, VA 22030, USA}

\author[0000-0001-6812-7938]{Tracy E. Clarke}
\affiliation{U.S. Naval Research Laboratory, 4555 Overlook Ave SW, Washington, DC 20375, USA}

\author[0000-0002-6454-861X]{Simona Giacintucci}
\affiliation{U.S. Naval Research Laboratory, 4555 Overlook Ave SW, Washington, DC 20375, USA}

\author[0000-0002-5187-7107]{Wendy Peters}
\affiliation{U.S. Naval Research Laboratory, 4555 Overlook Ave SW, Washington, DC 20375, USA}

\author[0000-0002-3984-4337]{Scott W. Randall}
\affiliation{Center for Astrophysics | Harvard \& Smithsonian, 60 Garden St., Cambridge, MA, 02138}

\author[0000-0002-0587-1660]{Reinout J. van Weeren}
\affiliation{Leiden Observatory, Leiden University, PO Box 9513, 2300 RA Leiden, The Netherlands}

\author[0000-0002-5222-1337]{Arnab Sarkar}
\affiliation{Kavli Institute for Astrophysics and Space Research, MIT, 70 Vassar St, Cambridge, MA 02139}

\author[0000-0001-5636-7213]{Lawrence Rudnick}
\affiliation{Minnesota Institute for Astrophysics, University of Minnesota. 116 Church St. SE, Minneapolis, MN 55455}

\author[0000-0002-0485-6573]{Elizabeth L. Blanton}
\affiliation{Institute for Astrophysical Research and Department of Astronomy, Boston University, Boston, MA 02215, USA}

\author[0000-0002-4462-0709]{Kyle Finner}
\affiliation{IPAC, California Institute of Technology, 1200 E California Blvd., Pasadena, CA 91125, USA}

\author[0000-0003-3816-5372]{Tony Mroczkowski}
\affiliation{European Southern Observatory, Karl-Schwarzschild-Straße 2, 85748 Garching bei München, Germany}

\author[0000-0003-0297-4493]{Paul Nulsen}
\affiliation{Center for Astrophysics | Harvard \& Smithsonian, 60 Garden St., Cambridge, MA, 02138}
\affiliation{ICRAR, University of Western Australia, 35 Stirling Hwy, Crawley, WA 6009, Australia}

\begin{abstract}
We present new radio observations of the galaxy cluster merger CIZA J0107.7+5408 (CIZA0107), a large, roughly equal mass, post-core passage, dissociative binary system at \texttt{z} = 0.1066. CIZA0107 is an elongated, disturbed system, hosting two subclusters with optical galaxy number density peaks offset from their associated X-ray density peaks and double-peaked diffuse radio structure. We present new 240-470 MHz and 2.0-4.0 GHz Very Large Array observations of CIZA0107. We image the diffuse emission at high resolution, constrain its integrated spectrum, and map the spectral index distribution. We confirm the presence of steep-spectrum ($\alpha \sim$ -1.3) emission on a scale of $\sim$0.5 Mpc in both subclusters. We identify two smaller ultra-steep spectrum (USS; $\alpha < -2$) regions, superimposed on larger-scale radio emission associated with the southwestern subcluster. At 340 MHz, we detect a radio edge bounding the emission to the south and show that it is coincident with a weak (M $\sim$ 1.2) shock identified in the \textit{Chandra} image. At 3 GHz, the emission does not show any corresponding edge-like feature, and in fact it extends beyond the shock. We investigate the nature of the emission in CIZA0107 and find that, while the system may host a double halo structure, we cannot rule out a scenario in which the emission arises from two relics projected on the central cluster regions.

\end{abstract}


\section{Introduction} \label{sec:intro}
Galaxy clusters are the largest virialized systems in the Universe, thought to form via gravitational collapse at the intersections of the large-scale structure filaments that make up the cosmic web \citep[][]{Kravtsov_2012}. A typical cluster has a mass of $10^{14-15} M_{\odot}$, of which only 5\% is from the component galaxies. Roughly 80\% of the mass is dark matter. A hot, diffuse intra-cluster medium (ICM) accounts for the remaining 15\% \citep[$10^{7-8}$ K, $\sim 10^{-3}$ cm$^{-1}$;][]{sarazin1988,bohringer2010}. Emission from the ICM is dominated by thermal bremsstrahlung and line emission and is strongest in the X-ray regime.

Clusters grow and evolve via a hierarchical process of mergers and accretion \citep[][]{Planelles_2009}. Cluster mergers are the most energetic events in the Universe since the Big Bang \citep[$\sim 10^{64}$ ergs on a few Gyr timescales][]{sarazin2002,tormen2004}. They reach collisional velocities of $>$ 2000 km s$^{-1}$ \cite[][]{sarazin2000}. A significant part of the colossal energy injected during the merger is dissipated through shocks, which heat and compress the ICM, enhancing the ICM's thermal emission in the X-ray regime \citep[][]{Markevitch_2007,brunettijones2014,forman1982,2022ApJ...935L..23S}. A small fraction of the energy injected during a merger is dissipated as turbulence in the ICM \citep[][]{brunettijones2014}. Clusters also contain non-thermal components (e.g., relativistic particles and $\mu$G magnetic fields) as evidenced by the existence of cluster-wide regions of diffuse synchrotron radio emission known as radio halos and relics \citep[][]{brunettijones2014,van_Weeren_2019,feretti2012}.

Radio halos are extended ($\geq$ 1 Mpc) diffuse radio sources typically located in cluster centers \citep[][]{brown2011}. They are generally unpolarized ($\leq$ 5\%), and follow the smooth surface brightness profile of the X-ray emitting gas. Halos are typically characterized by steep spectrum emission (where $\alpha <$ -1)\footnote{assuming $S_\nu$ $\propto$ $\nu^{\alpha}$, where $S$ is the flux density, $\nu$ is the frequency, and $\alpha$ is the spectral index}. Radio relics are non-thermal synchrotron structures typically found at the cluster peripheries which have irregular, elongated morphologies (see \cite{van_Weeren_2019} for a review). At GHz frequencies, they are found to be highly polarized, at $\sim$30\% \citep[][]{digennaro2021,ensslin1998relics}.

Radio halos are believed to originate via one of two scenarios. The first is the favored explanation, wherein seed relativistic electrons are re-accelerated \textit{in situ} via a Fermi-II process powered by the turbulence injected in the ICM by cluster mergers \citep[][]{brunetti2001reaccel,petrosian2001, brunetti2001ICscatter}. The necessary seed electrons can be produced via a combination of active galactic nuclei (AGN) activity, accretion shocks, and galactic outflows \citep[][]{brunettijones2014}. The second formation mechanism is a hadronic origin, wherein synchrotron-emitting electrons are produced as decay products of heavy nuclei/particle collisions: the long lifetimes of cosmic ray protons (CRps) allow them time to diffuse over the megaparsec scales exhibited by halos. Then, through collisions with the ICM heavy nuclei, they produce cosmic ray electrons (CRes) throughout the cluster environment, where they emit synchrotron radiation as they interact with the cluster magnetic field. \citep[][]{dennison1980,blasicolafrancesco1999,pfrommerensslin2004}.

Radio relics are thought to be driven by merger shock waves \citep[][]{vanweerenreview}, and evidence has been found for a spatial connection between radio relics and merger shocks detected in the X-ray regime as ICM density and temperature discontinuities (see \cite{botteon2016relic115,botteon2016elgordorelic,akamatsu2015}, and references therein). However, not all relics have been found to have an associated shock, possibly due to limitations in shock detection given the low X-ray counts in the cluster outskirts \citep[][]{ZuHone_2012}. 

Fermi-I diffuse shock acceleration (DSA) is accepted as the mechanism powering radio relics, wherein charged particles are accelerated to relativistic energies via scattering both upstream and downstream by magnetic inhomogeneities \citep[][]{fermi1949,blandfordeichler1987}. In the presence of the cluster-scale magnetic fields, they produce synchrotron emission, which is observable as diffuse, arc-like radio relics \citep[][]{brunettijones2014,van_Weeren_2019}. This emission follows a power law spectrum, where $-1.5 < \alpha < -1$ \citep[][]{blandfordeichler1987,rajpurohit2020}. The emission generates a spectral index profile in line with that observed in radio relics, which exhibit a flatter spectrum near the shock position \citep[][]{hoeftbruggen2007,kang2012}. DSA from the existing thermal pool cannot entirely explain the observed luminosities of relics; \cite{botteon2020} found that acceleration of the thermal pool via DSA in weak shocks with Mach numbers of $\mathcal{M}<$ 3-4 is insufficient to achieve the required efficiencies \citep[][]{vanweeren2017accel,botteon2020}. This tension can be resolved if radio relics are created by the re-acceleration of a pre-existing population of mildly relativistic CRes \citep[][]{markevitch2005,pinzke2013,capriolispitkovsky2014,dominguezfernandez2021}. There is both morphological and spectral evidence that the tails of radio galaxies and/or other AGN activity can provide this necessary seed electron population \citep[][]{bonafede2014,vanweeren2017accel,digennaro2018}.

A cluster merger can launch a pair of shock waves propagating in opposite directions along the collision axis on either side of the core \citep[][]{vanweeren2011relics}. These shocks can both result in particle acceleration, which can be observed as a cluster with a double relic morphology.

Some clusters have been found to exhibit both a radio halo and a distinct radio relic. In some cases, the components are less distinct, and the radio halos exhibit a sharp radio edge. This rare morphology is sometimes referred to as a radio halo-shock edge, as the radio edge appears to trace a shock front in the ICM \citep[][]{van_Weeren_2019,brown2011,shimwell2014,vanweeren2016,wang2018,botteon2024}. The nature of these structures is still unclear \citep[see][]{markevitch520}. One possibility is that the radio edge is driven by shock-induced adiabatic compression of pre-existing CRes. Another possibility is that the edge arises from seed electrons that are re-accelerated at the shock front. As the electrons move downstream, they are further re-accelerated by turbulence in the merger, leading to the formation of a radio halo. In this case, it is possible that the two features may blend, forming a radio halo-shock edge. These systems are not well parameterized; no polarized emission has been observed \citep[][]{shimwell2014}, and no clear strong downstream spectral gradients due to electron energy losses have been identified \citep[][]{vanweeren2016,rajpurohit2018}.

Finally, clusters can also host extended radio sources known as radio phoenixes. These are diffuse radio objects that display a variety of morphologies, usually highly irregular, filamentary and polarized \citep[][]{kale2012,mandal2020}. They are found to be at most hundreds of kpc in length and located between the cluster center and periphery \citep[][]{degasperin2015,van_Weeren_2019}. Phoenixes are thought to trace fossil AGN plasma that has been re-energized; for example, adiabatically compressed via the passage of an ICM shock \citep[][]{kempner2004}. This scenario also results in the very steep spectrum typically observed for phoenixes, driven by synchrotron losses \citep[$\alpha <$ -1.5;][]{vanweeren2011a}.

\subsection{CIZA J0107.7+5408} \label{subsec:ciza}

CIZA J0107.7+5408, hereafter CIZA0107, is a nearby, post-core passage, binary cluster merger, with a redshift of z = 0.107 \citep[][]{ebeling2002}. It has been previously identified as ZwCl 0104.9+5350 \citep[]{vanweeren2011}. CIZA0107 is part of the X-ray survey for Clusters in the Zone of Avoidance, which was the first statistically complete catalog of X-ray selected galaxy clusters behind the Galactic plane \citep[CIZA;][]{kocevski2003clusters}. 

CIZA0107 hosts two optical density peaks, with associated but offset X-ray emission peaks. As two clusters collide, their ICM components experience ram pressure forces and slow down, while the effectively collisionless component galaxies and dark matter halos do not. The effect is significant enough to displace the ICM X-ray peaks from the peaks of the optical galaxy density and dark matter, where the locations of the dark matter peaks are identified via weak lensing \citep[for a review, see][]{umetsu2020}. The rare systems that display this offset, such as CIZA0107, are called dissociative galaxy cluster mergers, less than ten of which have been confirmed (see \cite{mcdonald2022} and references therein).

\cite{Randall_2016}, hereafter referred to as R16, presented X-ray, optical, and radio results for CIZA0107, illustrating the double-peaked morphology, and identifying the brightest cluster galaxies (BCGs) at 01:07:55.959, +54:10:13.68 and 01:07:40.697, +54:06:31.94. They report ICM X-ray emission with a clear elongation from NE to SW, which they interpret as evidence of a dynamically disturbed merging system. A thorough analysis of the weak lensing mass distribution for CIZA0107 was presented in \cite{Finner_2023}. The weak lensing analysis also finds evidence for two subclusters. The NE subcluster was found to be compact and round, while the SW subcluster was found to be elongated in both the west-east and NE to SW directions. These results were interpreted as a possible sign of substructure below the detection threshold of the optical data. 

CIZA0107 has been previously observed with the Giant Metrewave Radio Telescope at 240 and 610 MHz (R16) and by the Very Large Array (VLA) at 73.8 MHz as part of the VLA Low-frequency Sky Survey redux \cite[VLSSr;][]{lane2014}. These observations revealed diffuse emission classified as a giant radio halo by \cite{vanweeren2011} and separately as two radio relics with a N-S extension by R16. R16 further identified two distinct peaks associated with the SW subcluster that exhibit ultra-steep spectrum (USS) emission ($\alpha \sim$ -2). The USS sources have spectra consistent with an old non-thermal electron population that has been re-energized via adiabatic compression by the passage of an ICM shock \cite[such systems are commonly referred to as a radio phoenix;][]{ensslin2001}.

In this paper, we present new Karl G. Jansky Very Large Array (VLA) observations of CIZA0107 at 340 MHz and 3 GHz. We compared this data to existing \textit{Chandra} X-ray observations in order to study the cluster morphology. Unless otherwise noted, a flat $\Lambda$CDM cosmology is adopted, with a $\Omega_{\Lambda}$ = 0.69, $\Omega_{m}$ = 0.31, and $H_{0}$ = 67.7 km s$^{-1}$ Mpc$^{-1}$ \cite[][]{planck2020}. Scale at the redshift of CIZA0107 is 1.95 kiloparsecs per arcsecond.

\section{Observations} \label{sec:obs}

We observed CIZA0107 at 340 MHz and 3 GHz, utilizing the VLA under Project Code SK0442. All four configurations were used in order to provide matched resolutions for both the compact sources and the diffuse emission. Details of the observations are summarized in Table \ref{tab:JVLAobs}.

\begin{deluxetable*}{cccccccc}
\tablenum{1}
\tablecaption{VLA Observations}
\tablewidth{0pt}
\tablehead{
\colhead{Band/Config} & \colhead{ Date} & \colhead{Frequency} & \colhead{Amplitude Cal} &
\colhead{Gain Cal} & \colhead{$T_{int}$} & \colhead{SPW} & \colhead{Channels}\\
\colhead{} & \colhead{} & \colhead{MHz} & \colhead{} &
\colhead{} & \colhead{hr} & \colhead{} & \colhead{per SPW} }
\startdata
P/A & 5 Sept, 2019 & 224-480 & 3C48 & - & 4.5 & 16 & 128\\
P/B & 9-10 Mar, 2019 & 224-480 & 3C48 & - & 4.75 & 16 & 128 \\
S/C & 16 Mar, 2020 & 2000-4000 & 3C48 & J2355+4950 & 4.25 & 16 & 64\\
S/D & 17-18 Jan, 2020 & 2000-4000 & 3C48 & J2355+4950 & 4.25 & 16 & 64\\
\enddata
\caption{\textbf{Notes}: Column 1 - VLA observation band and telescope configuration. Column 2 - Observation date. Column 3 - Frequency coverage in MHz. Column 4 - VLA flux density calibrator(s). Column 5 - VLA complex gain calibrator(s). Column 6 - Integration time, rounded to the nearest quarter hour. Column 7 - Number of spectral windows, pre-processing. Column 8 - Number of channels per spectral window, pre-processing.}
\label{tab:JVLAobs}
\end{deluxetable*}

\subsection{\textit{P}-band (230-470 MHz) VLA Observations} \label{subsec:pband}

High sensitivity radio observations of CIZA0107 were made with the VLA at 224-480 MHz (P-band) with the telescope in A and B configurations. This provides a resolution of $\sim$ 5$\arcsec{}$, with a maximum recoverable scale of $\sim$3$^{\prime}$ at A configuration, and a resolution of $\sim$15$\arcsec{}$, with a maximum recoverable scale of $\sim$10$^{\prime}$ in B configuration.  Details of the observations and calibrators used are given in Table \ref{tab:JVLAobs}.
        
\subsection{\textit{S}-band (2-4 GHz) VLA Observations} \label{subsec:sband} 

Observations were made of CIZA0107 at 2-4 GHz (S-band), with the VLA in C and D configurations. This provides a resolution of $\sim$7$\arcsec{}$ in C configuration. We note that the shortest baselines in C configuration are provided by a single antenna, and while these correspond to a maximum recoverable scale of $\sim$4$^{\prime}$, the configuration has considerably less sensitivity to diffuse emission. D configuration provides a resolution of $\sim$23$\arcsec{}$, with a maximum recoverable scale of $\sim$8$^{\prime}$.

\section{Data Reduction} \label{sec:data}

The flux densities for the primary calibrations in all configurations were taken from the \cite{perley2017} extension to the \cite{baars1977} scale. To refine the gain calibration, several rounds of self-calibration were performed for all observations, beginning with phase-only and proceeding to one round of amplitude and phase.

\subsection{A Configuration}
\label{subsec:aconfig}
The A configuration observations with a central frequency of 340 MHz were reduced and calibrated with Obit \cite[][]{obit}, using tasks in the standard \texttt{EVLALowBandPipe.py} pipeline. Basic calibration steps include an initial flagging for poor instrumental response and/or saturation issues from radio frequency interference (RFI), as well as known narrow-line RFI. The initial 14 and last eight frequency channels, where the sensitivity diminishes due to the passband, were removed, as well as the initial six seconds of each scan. The calibrator data were then filtered in time and frequency to remove any remaining RFI. Corrections for the system power were then applied, and observations of the calibrator were calibrated for complex delay and frequency-dependent complex gains. These were applied to the calibrator data before the editing for RFI was repeated. The calibration solutions were re-calculated, and then smoothed and applied to the target data. The calibrated target data were then filtered for RFI using standard tasks.  

No observations of a secondary complex gain calibrator were available to correct for ionospheric contribution to the line-of-sight phases. Instead, phase-only corrections were calculated by a direction-independent calibration of the target data to a sky model based on strong sources from the NRAO VLA Sky Survey \citep[NVSS;][]{condon1998}, scaled by a spectral index of $\alpha=-0.75$. This removes any positional offsets, providing a final positional accuracy that should be at least comparable to the arcsecond precision of NVSS. Any leftover RFI was removed using NRAO's Common Astronomy Software Applications \cite[CASA;][]{casanew2022} tasks \textsc{RFlag} and \textsc{TFCrop}. The final bandwidth was 225-465 MHz, with 7 spectral windows. 

\subsection{B Configuration}
\label{subsec:bconfig}

The B configuration observations with a central frequency of 340 MHz were reduced and calibrated manually with CASA, following standard procedures \citep[][]{CASA_PBandGuide}. RFI was removed with the \textsc{RFlag} and \textsc{TFCrop} functions. The final bandwidth was 272-416 MHz, with 5 spectral windows.

\subsection{C Configuration}
\label{subsec:cconfig}

The C configuration observations with a central frequency of 3 GHz were reduced and calibrated using the VLA pipeline (version 5.6.2) in CASA. Any remaining RFI was manually excised. The final bandwidth was 1988-3500 MHz, with 10 spectral windows.

\subsection{D Configuration}
\label{subsec:dconfig}

The D configuration observations with a central frequency of 3 GHz were reduced using the VLA pipeline (version 5.6.2) in CASA. Any remaining RFI was manually excised. The final bandwidth was 1988-3884 MHz, with 9 spectral windows.

\subsection{Point Source Subtraction} \label{subsec:sub}

In order to fully constrain the diffuse emission of the cluster merger, it was necessary to reduce the effects of the contaminant compact emission. Six compact sources were subtracted off at 340 MHz from within the 3$\sigma$ radio emission contours. Forty-five compact sources were subtracted off at 3 GHz from within the 3$\sigma$ radio emission contours. At both frequencies, a number of sources outside the $3\sigma$ contours were also subtracted. This process is discussed in detail in Appendix A.

\subsection{Imaging} \label{subsec:imaging}

Once calibration was complete, all final imaging was carried out with the W-stacking clean algorithm \citep[WSClean;][]{wsclean} to account for the widefield errors. The cell size was taken to be a fifth of the restoring beam. Multiple resolution scales were used at 0, 3 $\times$ resolving beam, 5 $\times$ resolving beam \citep[][]{cornwell2008}. The parameter \textsc{-fit-spectral-pol} in WSClean was used to model the frequency dependence of the sky, and to account for the VLA's wideband receivers.

Clean masks for compact emission were employed at all stages, and were generated using the Python Blob Detector and Source Finder \cite[PyBDSF;][]{pybdsf} source detection package. The final images were corrected for the primary beam attenuation using the beam models provided by the Astronomical Image Processing System \cite[AIPS;][]{aipsgreisen}. Final image properties are described in Table \ref{tab:ciza_improps}.

\begin{figure*}[ht!]
\centering
\begin{minipage}{0.45\textwidth}
    \includegraphics[width=\linewidth]{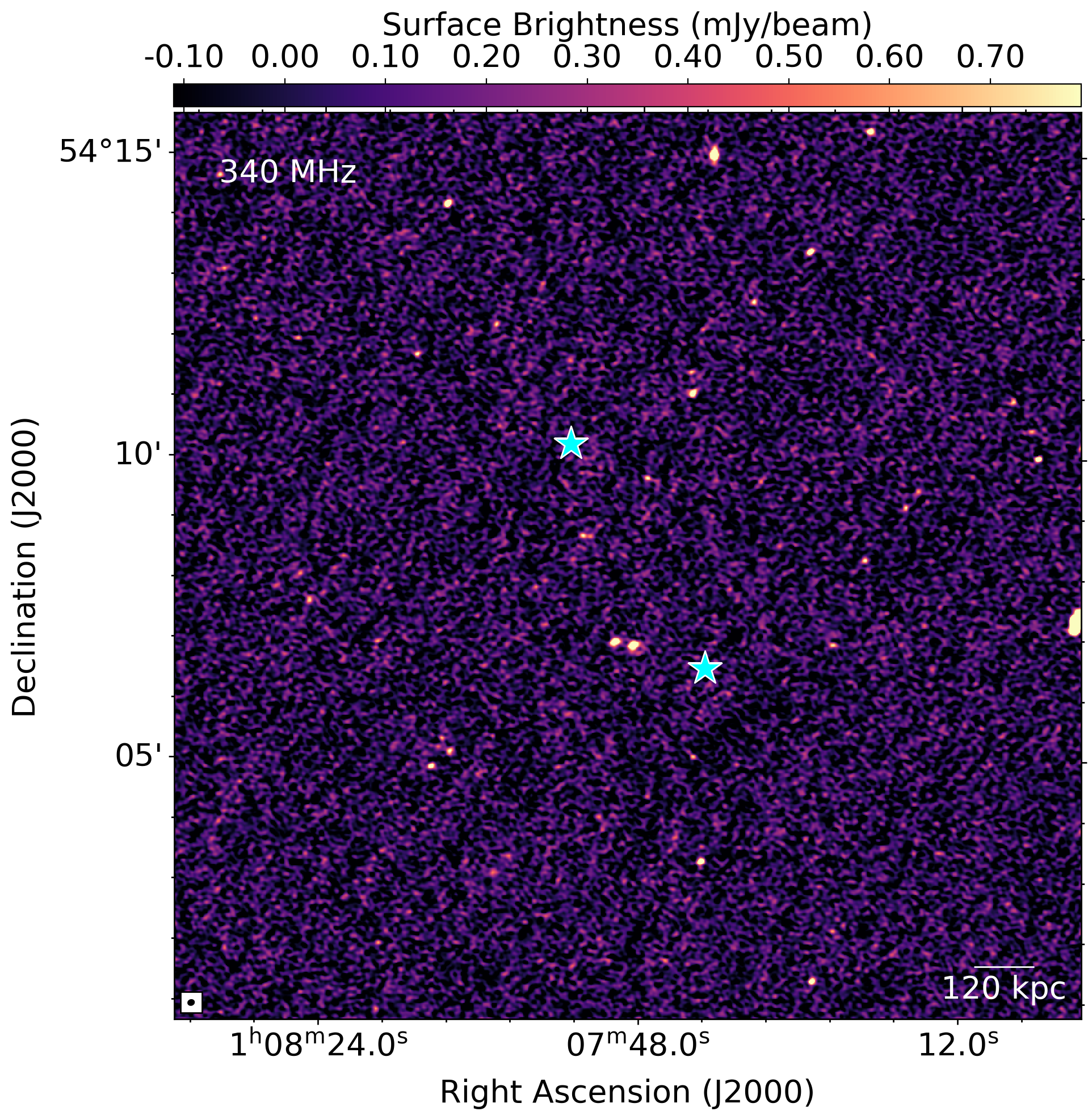}
\end{minipage}
\begin{minipage}{0.45\textwidth}
    \includegraphics[width=\linewidth]{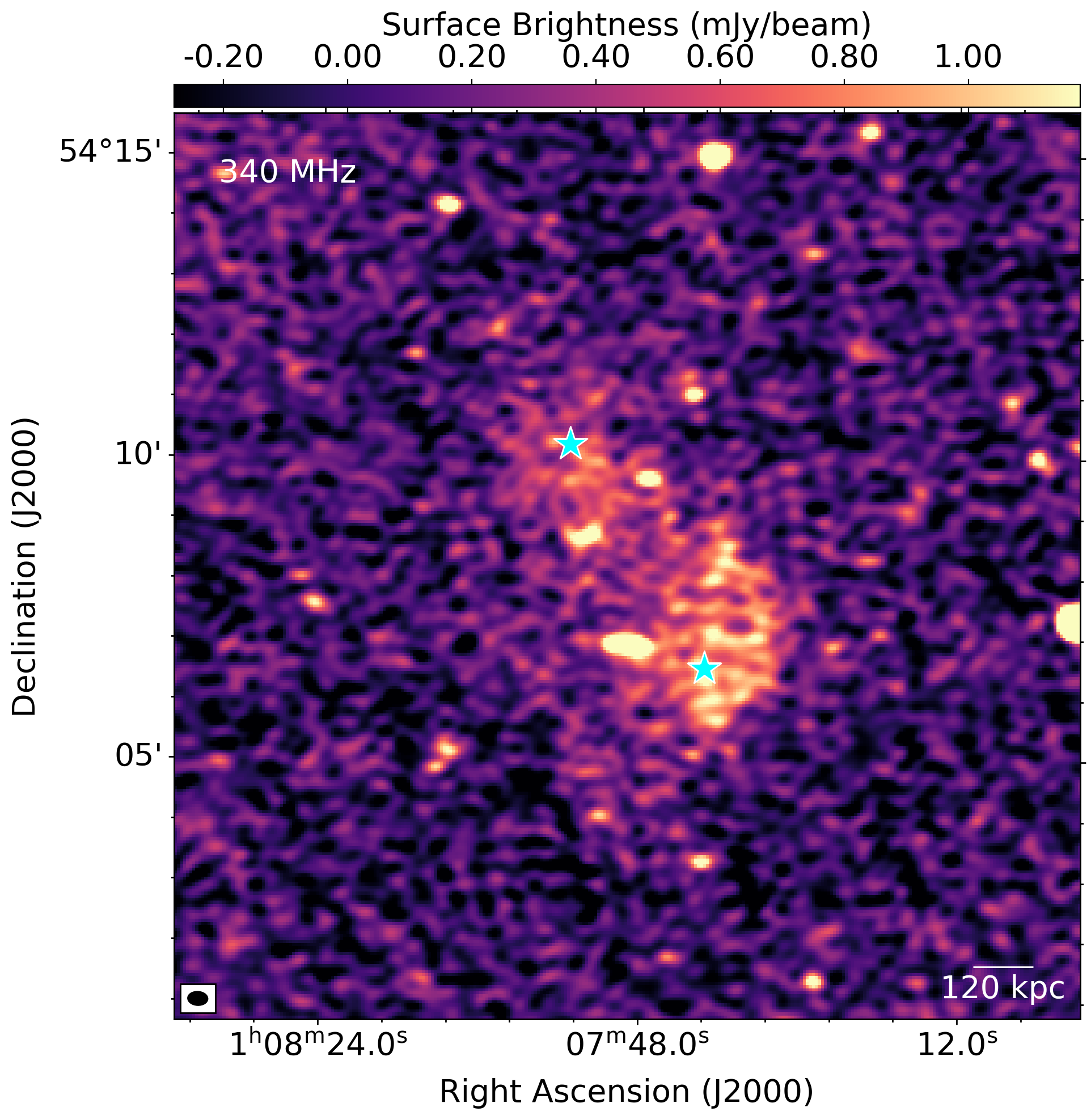}
\end{minipage}
\begin{minipage}{0.45\textwidth}
    \includegraphics[width=\linewidth]{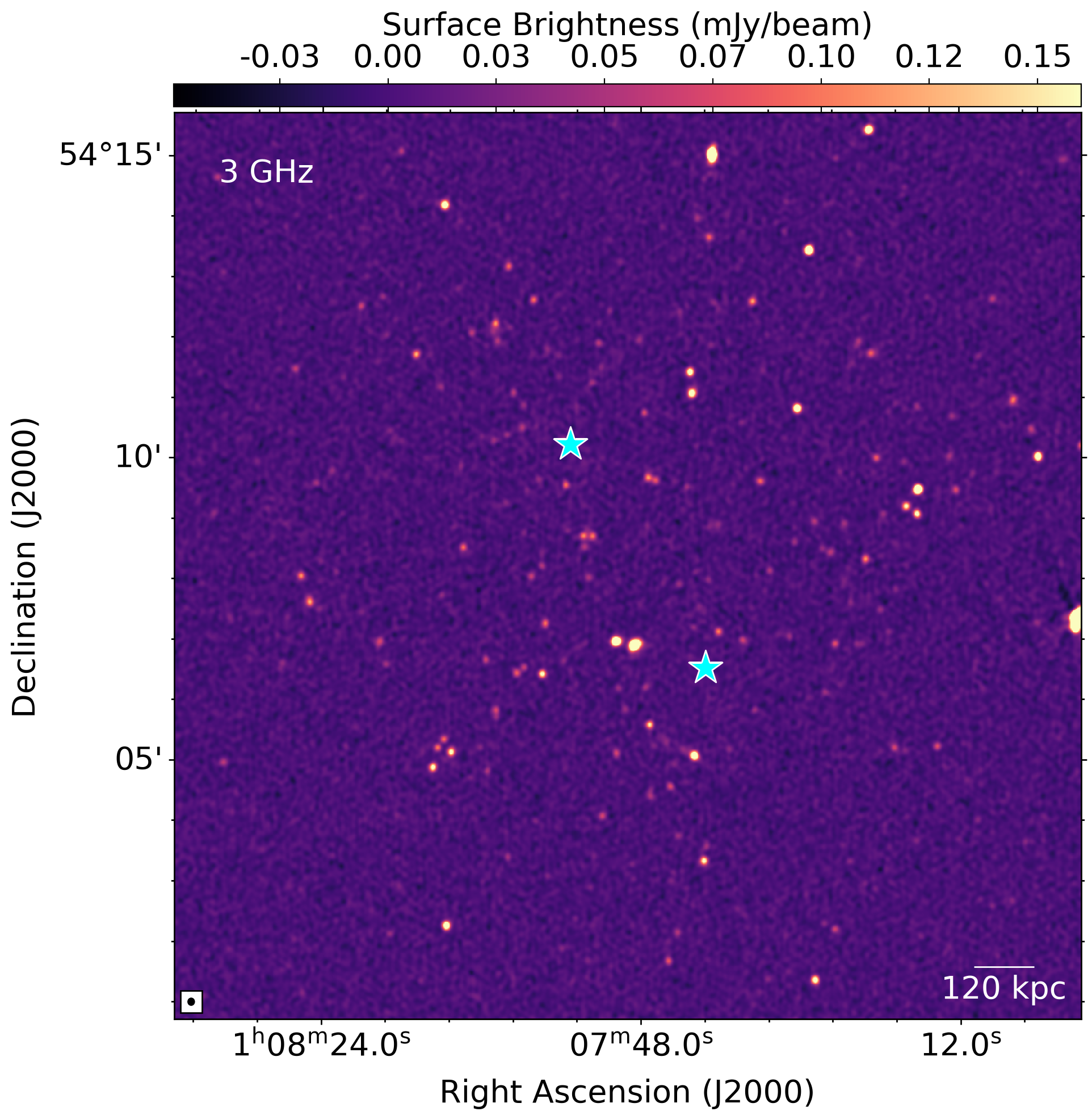}
\end{minipage}
\begin{minipage}{0.45\textwidth}
    \includegraphics[width=\linewidth]{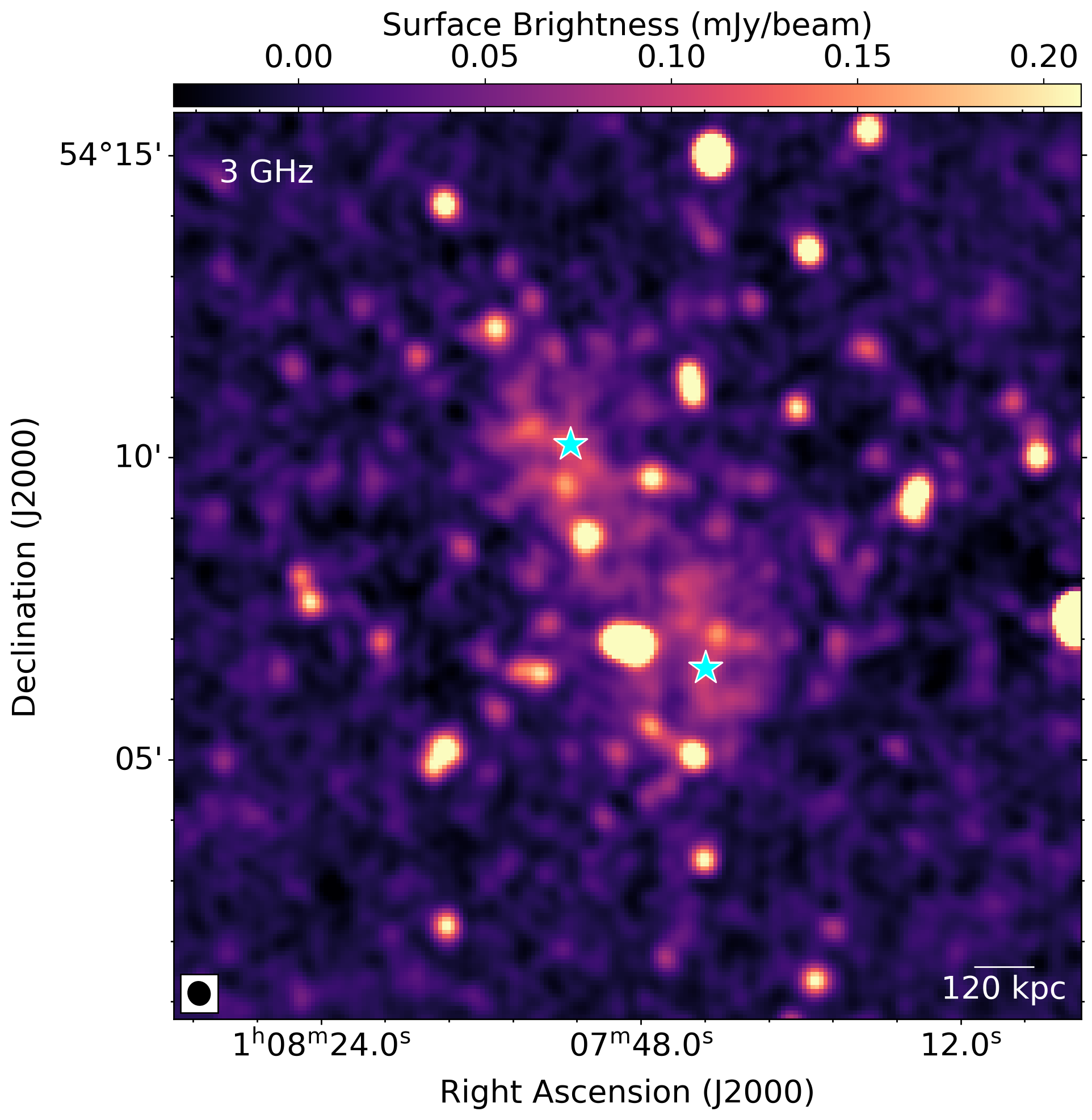}
\end{minipage}
\caption{A (top left): VLA 340 MHz A-configuration image with a 5.93'' $\times{}$ 5.01'' $\times{}$ -70$^{\circ}$ beam. B (top right): VLA 340 MHz B-configuration image with a 19.44'' $\times{}$ 13.28'' $\times{}$ 86$^{\circ}$ beam. C (bottom left): VLA 3 GHz C-configuration image with a 6.65'' $\times{}$ 6.10'' $\times{}$ -1$^{\circ}$ beam. D (bottom right): VLA 3 GHz D-configuration image with a 23.33'' $\times{}$ 21.46'' $\times{}$ 20$^{\circ}$ beam. Cyan stars mark the location of the subcluster BCGs. The physical scale is 1.95 kiloparsecs per arcsecond.}
\label{fig:ciza_nativeims}
\end{figure*}

\begin{deluxetable*}{cccccc}
\tablenum{2}
\tablecaption{Image Properties\label{tab:ciza_improps}}
\tablewidth{0pt}
\tablehead{
\colhead{Image} & \colhead{Figure} & \colhead{Central Frequency} & \colhead{Resolution} & \colhead{\textit{uv}-taper} & \colhead{$\sigma_{rms}$} \\
\colhead{} & \colhead{} & \colhead{MHz} & \colhead{'','',$^{\circ}$} & \colhead{''} & \colhead{$\mu Jy$/beam}
}
\startdata
1 & \ref{fig:ciza_nativeims}A & 340/A & 5.93, 5.01, -70 & - & 213 \\
2 & \ref{fig:ciza_nativeims}B & 340/B & 19.44, 13.28, 86 & - & 177 \\
3 & \ref{fig:ciza_nativeims}C & 3000/C & 6.65, 6.10, -1 & - & 4.87 \\
4 & \ref{fig:ciza_nativeims}D & 3000/D & 23.33, 21.46, 20 & - & 7.95 \\
5$\dagger$ & \ref{fig:ciza_subims}A & 340/B & 18.98, 13.01, 87 & - & 167 \\
6$\dagger$ & \ref{fig:ciza_subims}B & 3000/D & 23.13, 21.41, 18 & - & 7.41 \\
7$\dagger$ & \ref{fig:ciza_subims}C & 340/B & 40, 40, 0 & 30'' & 397 \\
8$\dagger$ & \ref{fig:ciza_subims}D & 3000/D & 40, 40, 0 & 30'' & 17.1 \\
\enddata
\caption{\textbf{Notes}: Column 1 - Image number, referenced throughout the text; $\dagger$ indicates and image that has been point source subtracted. All images used in final analysis. Column 2 - Figure associated with each image number. Column 3 - Central frequency in MHz of observations shown in each image, as well as telescope configuration. Column 4 - Restoring beam: major axis, minor axis in arcseconds, position angle in degrees measured east of north. Column 5 - \textit{uv}-taper used to generate image in arcseconds, if applicable. Column 6 - 1-$\sigma$ sensitivity, measured near the phase center.}
\end{deluxetable*}

Once point-source subtraction was complete, the observations at 340 MHz in the B configuration and at 3 GHz in the D configuration were tapered with a Gaussian taper to produce a resolution close to 40'' and then convolved with a 40'' circular beam to best display the extended emission. The final images parameters are described in Table \ref{tab:ciza_improps}, where the point-source subtracted images are designated with a dagger symbol. Imaging followed the same process as described above. In order to best trace the full extent of the diffuse emission, clean masks for point-source subtracted images were drawn, and employed at all stages.

\section{Radio Results} \label{sec:radioresults}

\subsection{Continuum Imaging} \label{subsec:radiocontinuum}

The new VLA 340 MHz and 3 GHz observations are shown in Figure \ref{fig:ciza_nativeims}. The different combinations of configuration and frequency yield different views of the diffuse radio emission morphology in the cluster field. In A configuration, at 340 MHz, only emission from compact sources is visible. In B-configuration, at 340 MHz, while still contaminated by compact emission, the radio emission associated with both subclusters is clearly seen as diffuse emission with a peak to the northeast and a second peak to the southwest. 

In C configuration, at 3 GHz, only compact emission is again visible. In D configuration, at 3 GHz, the diffuse emission is once again visible, though also contaminated by compact emission. As seen in the P-band, B configuration image, there are two peaks in the radio surface brightness, one to the northeast, and a second to the southwest. 

\begin{figure*}[ht!]
\centering
\begin{minipage}{0.45\textwidth}
    \includegraphics[width=\linewidth]{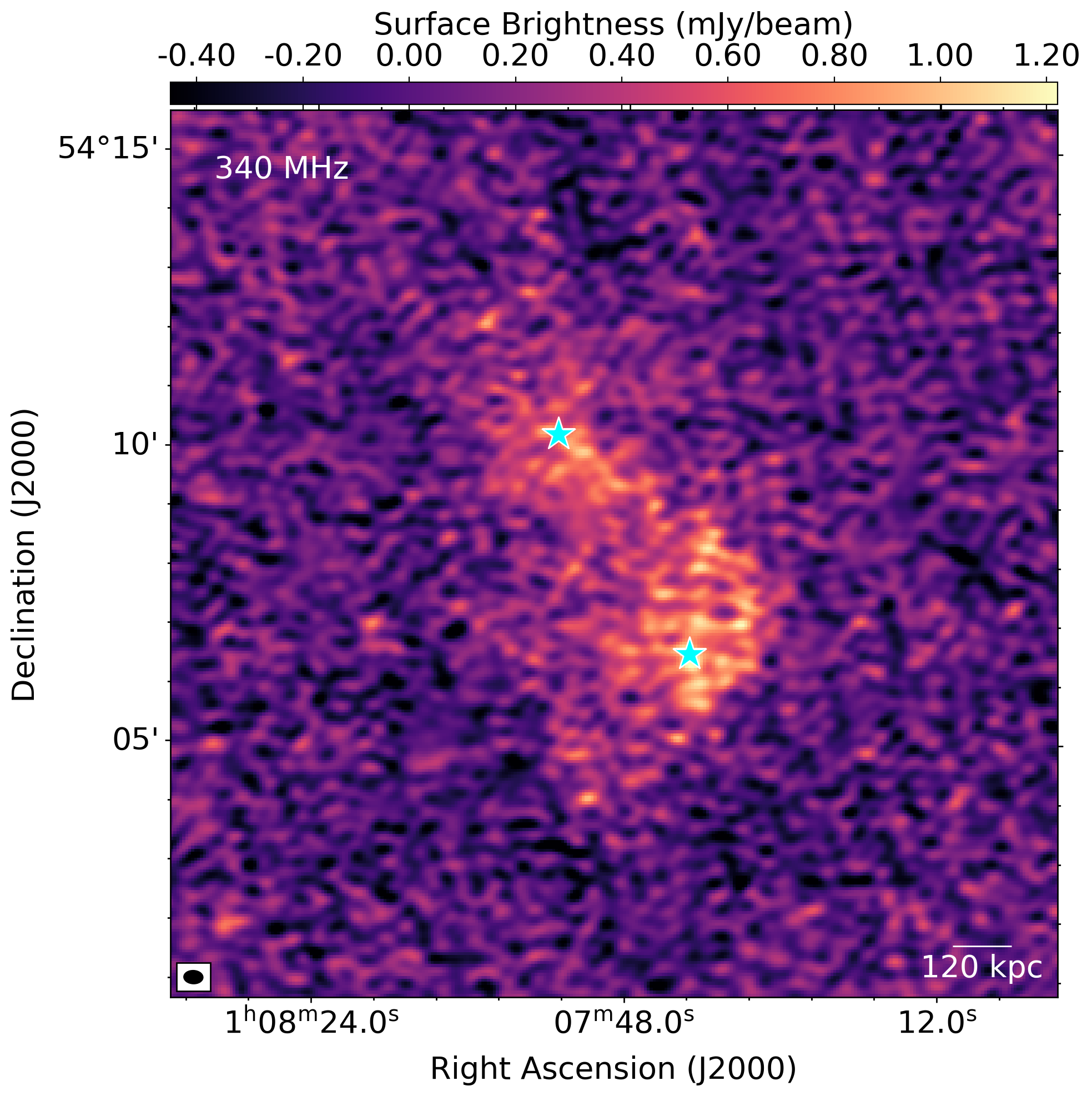}
\end{minipage}
\begin{minipage}{0.45\textwidth}
    \includegraphics[width=\linewidth]{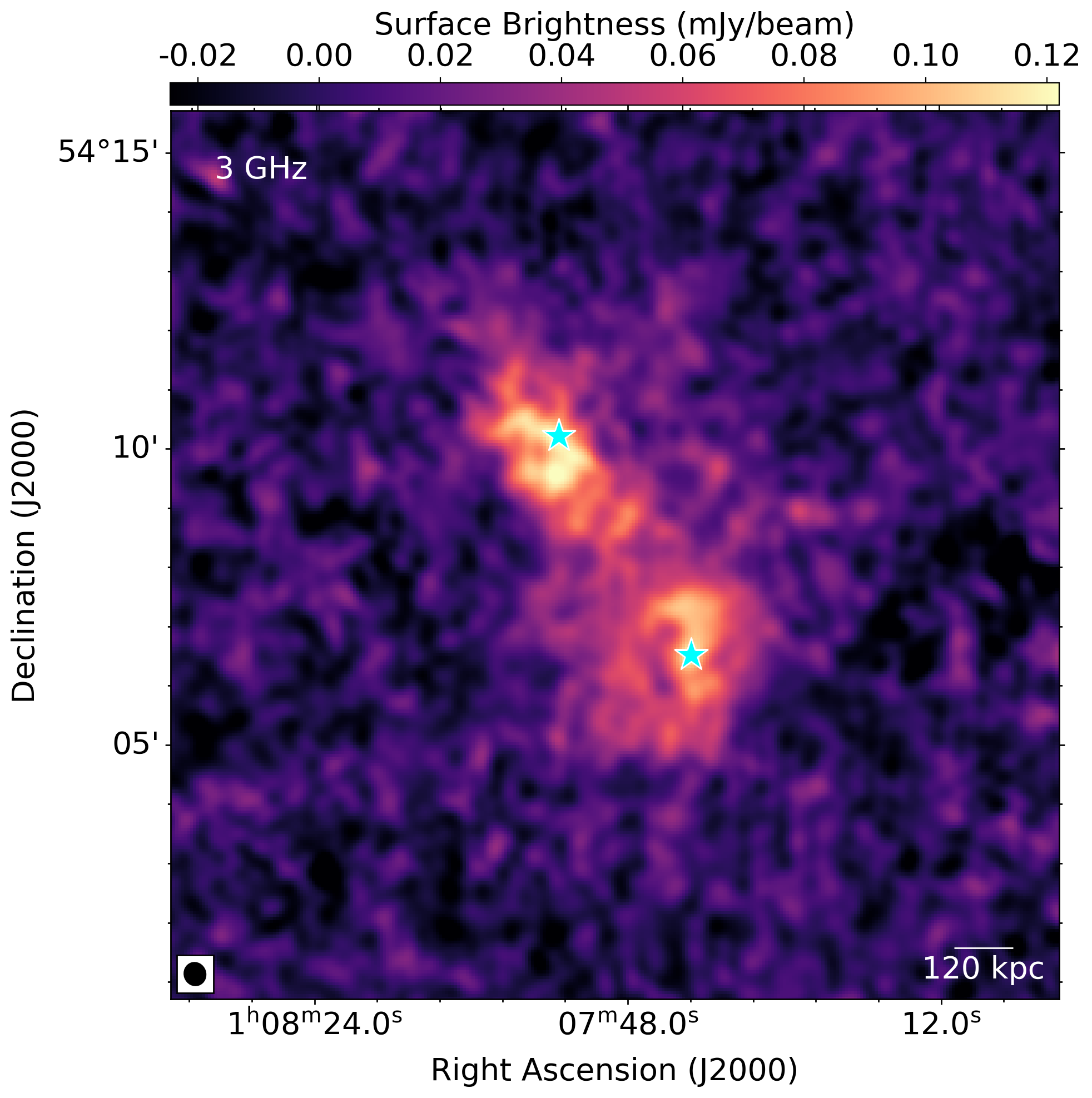}
\end{minipage}
\begin{minipage}{0.45\textwidth}
    \includegraphics[width=\linewidth]{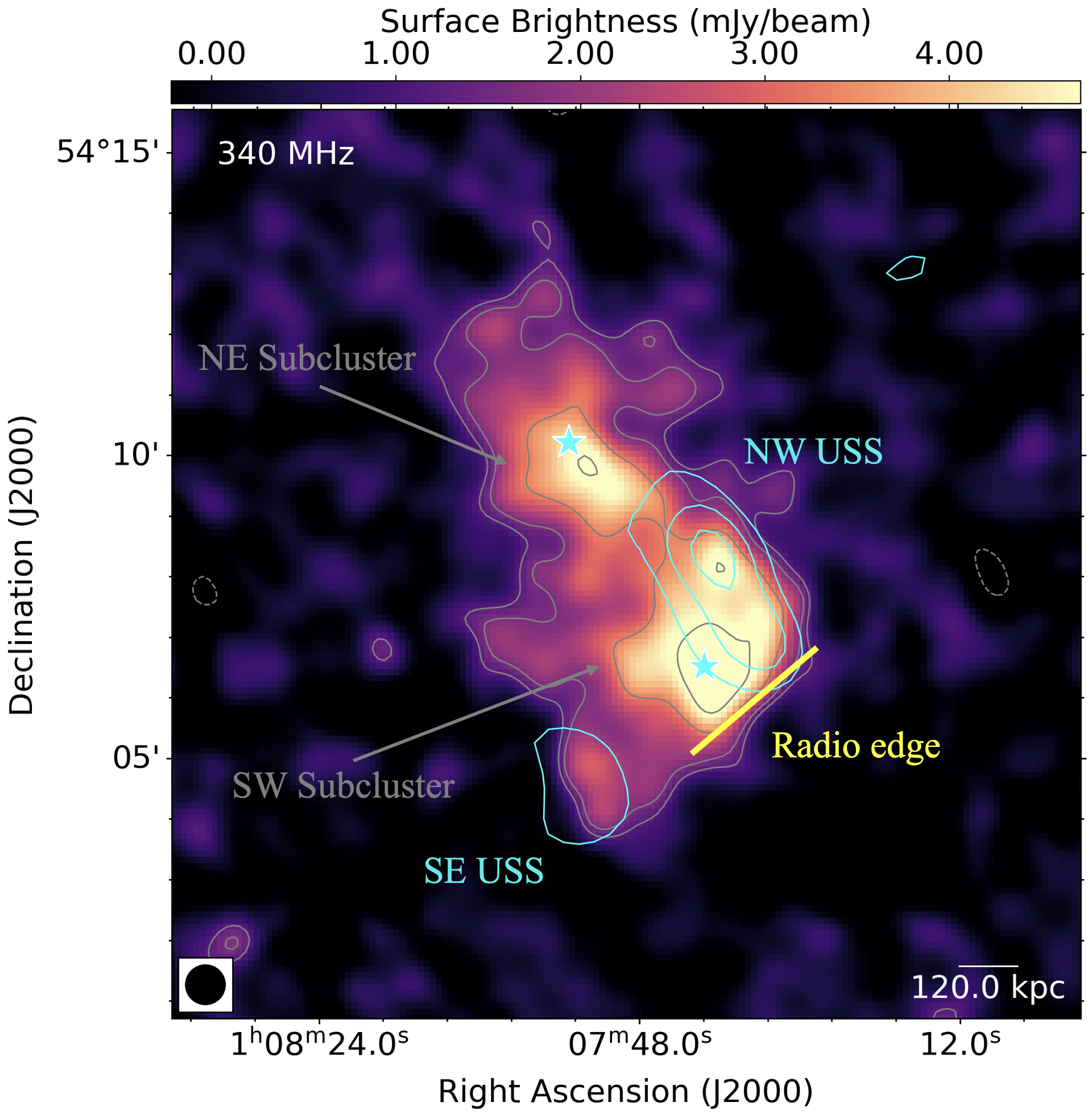}
\end{minipage}
\begin{minipage}{0.45\textwidth}
    \includegraphics[width=\linewidth]{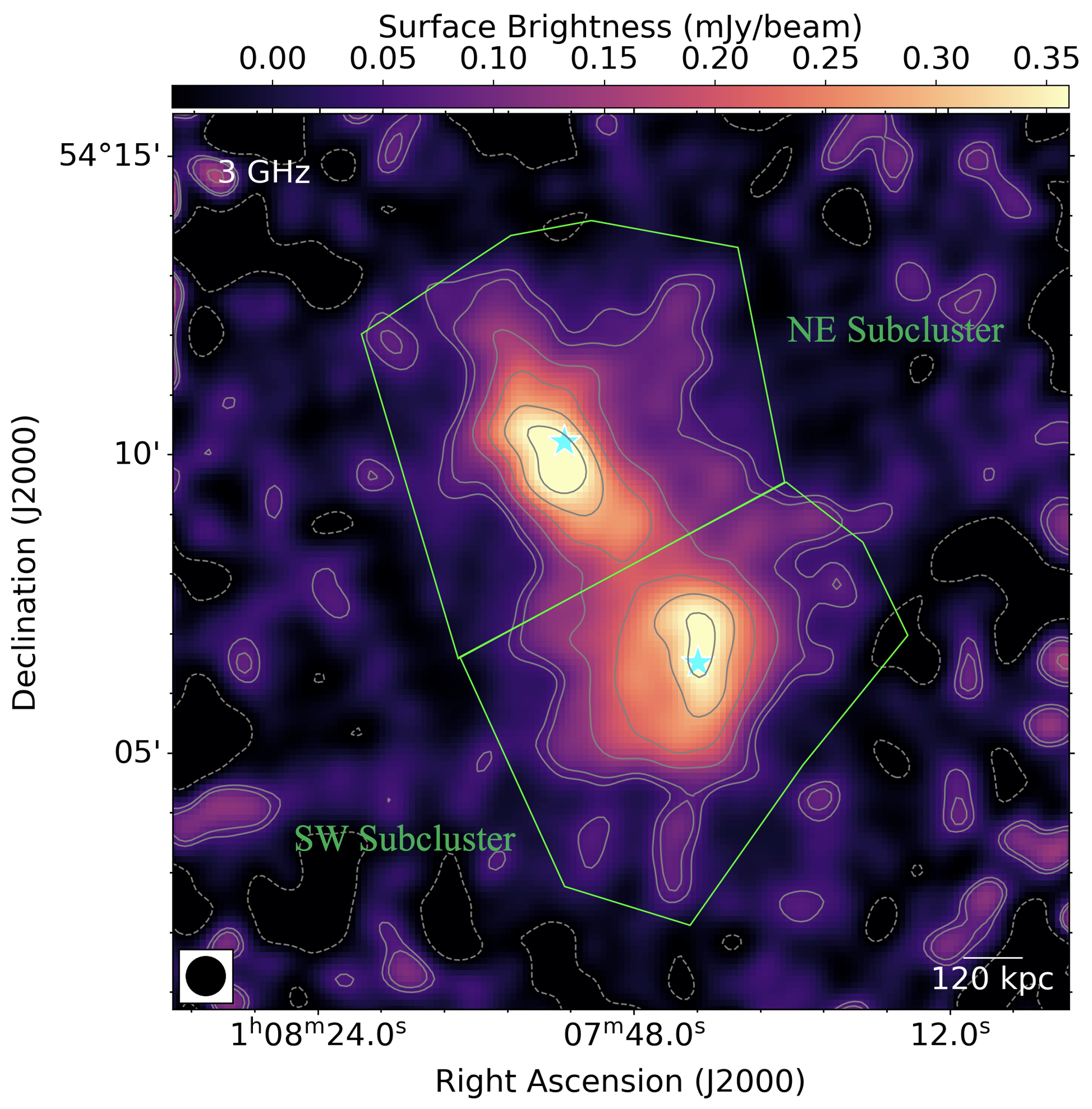}
\end{minipage}\caption{A (top left): VLA 340 MHz B-configuration image with compact emission subtracted out, and a 18.98'' $\times{}$ 13.01'' $\times{}$ 87$^{\circ}$ beam. B (top right): VLA 3 GHz D-configuration image with compact emission subtracted out, and a 23.13'' $\times{}$ 21.41'' $\times{}$ 18$^{\circ}$ beam. C (bottom left): VLA 340 MHz B-configuration image with compact emission subtracted out, convolved to a 40'' beam. Cyan contours trace the 74 MHz VLSSr emission, beginning at 3$\sigma$ and proceeding in integer multiples of $\sqrt{2}$, highlighting the ultra-steep spectrum regions. The sharp radio edge in the southwest is highlighted with a yellow line, the northeast and southwest subclusters are labeled, and the southeast and northwest ultra-steep spectrum regions are labeled. D (bottom right): VLA 3 GHz D-configuration image with compact emission subtracted out, convolved to a 40'' beam. The green boundaries mark the regions used for integrated flux extraction, and are labeled by subcluster. Both 40'' images are shown with gray contours beginning at 3$\sigma$ (3 GHz: 30.7 $\mu$Jy, 340 MHz: 431 $\mu$Jy) and proceeding in integer multiples of $\sqrt2$. Dashed grey contours trace the -3$\sigma$ emission. Cyan stars mark the locations of the BCGs. The physical scale is 1.95 kiloparsecs per arcsecond.}
\label{fig:ciza_subims}
\end{figure*}

The new VLA 340 MHz and 3 GHz observations with the compact sources removed as described in Sections \ref{subsec:sub} and \ref{sec:appendixA} are shown for the B and D configurations in Figures \ref{fig:ciza_subims}A and \ref{fig:ciza_subims}B at their native resolutions. The same 340 MHz and 3 GHz images are shown in Figures \ref{fig:ciza_subims}C and \ref{fig:ciza_subims}D, where they have been Gaussian tapered and convolved to a 40'' resolution. The locations of the BCGs are marked with cyan stars. Figure \ref{fig:ciza_subims}C also presents the nomenclature used in this paper.

The diffuse radio emission tracing the NE and SW subclusters is clear at both frequencies. We note a drop-off in the surface brightness at 340 MHz, labeled `radio edge' in Figure \ref{fig:ciza_subims}C, also visible in the non-point source subtracted image in Figure \ref{fig:ciza_nativeims}B. This feature is further discussed in Section \ref{sec:xray},
where we investigate its possible association with a shock front, and Section \ref{subsubsec:southwest}.

In Figure~\ref{fig:ciza_subims}C, the cyan contours trace the 74 MHz VLSSr regions identified by R16 as ultra-steep spectrum emission. The 74 MHz contours show two distinct peaks to the northwest and southeast of the diffuse emission associated with the southern subcluster. The region to the northwest is more elongated and brighter in comparison to the southeastern peak. Notably, they are aligned near-perpendicular to the merger axis. The northwestern VLSSr region is coincident with the brightest emission at 340 MHz while we see no evidence for excess emission at the location of USS regions in the 3 GHz image (Figure~\ref{fig:ciza_subims}D).

Angular and projected physical sizes of the diffuse emission associated with each subcluster and the overall merging system are presented in Table \ref{tab:subclusterprops}. The subcluster and full cluster merger extents are measured along the axis of the merger, within the 3$\sigma$ limits of the 40'' resolution images in Figures \ref{fig:ciza_subims}C and \ref{fig:ciza_subims}D.

The size of the USS regions, measured along the merger axis within the 3$\sigma$ contours at 74 MHz, is presented in Table \ref{tab:ussclusterprops}. 

\begin{deluxetable*}{cccccccc}
\tablenum{3}
\tablecaption{Cluster Radio Properties}
\tablewidth{0pt}
\tablehead{
\colhead{} & \colhead{$S_{340 MHz}$} & \colhead{log($L_{340 MHz}$)} & \colhead{$S_{3 GHz}$} & \colhead{log($L_{3 GHz}$)} & \colhead{$\alpha_{340}^{3000}$} & \colhead{$\theta(r_p)_{340  MHz}$} & \colhead{$\theta(r_p)_{3  GHz}$}\\
\colhead{} & \colhead{mJy} & \colhead{$W Hz^{-1}$} & \colhead{mJy} & \colhead{$W Hz^{-1}$} & \colhead{} &
\colhead{''(kpc)} & \colhead{''(kpc)}}
\startdata
NE & 113$\pm$12 & 24.52 & 7.37$\pm$0.39 & 23.36 & -1.27$\pm$0.05 & 305(620) & 340(680)\\
SW & 117$\pm$12 & 24.56 & 7.92$\pm$0.42 & 23.39 & -1.25$\pm$0.06 & 220(440) & 310(630) \\
Total & 231$\pm$17 & 24.85 & 15.2$\pm$0.58 & 23.67 & -1.25$\pm$0.05 & 520(1040) & 660(1340) \\
\enddata
\caption{\textbf{Notes}: Column 1 - Subcluster. Column 2 - Integrated flux $\pm$ error in mJy at 340 MHz. Column 3 - 340 MHz luminosity in $W Hz^{-1}$. Column 4 - Integrated flux $\pm$ error in mJy at 3 GHz. Column 5 - 3 GHz luminosity in $W Hz^{-1}$. Column 6 - Spectral index calculated between 340 MHz and 3 GHz. Column 7 - Angular and physical size, reported in arcseconds and kiloparsecs, as measured at 340 MHz along the axis of the merger. Column 8 - Angular and physical size, reported in arcseconds and kiloparsecs, as measured at 3 GHz along the axis of the merger.}
\label{tab:subclusterprops}
\end{deluxetable*}

\subsection{Spectral Analysis} \label{subsec:spixresults}

Integrated radio fluxes are presented in Table \ref{tab:subclusterprops}. These were measured separately for the radio emission associated with the northern and southern subclusters, and for the cluster merger as a whole. Figure \ref{fig:ciza_subims}D presents the regions used to measure the integrated flux density for the two separate subclusters. The two regions are shown on the 340 MHz image. Final integrated flux errors take into account flux scale uncertainties of 10\% at 340 MHz and 5\% at 3 GHz \citep[][]{NRAO_FluxDensityScale}, as well as image noise.

For the total emission associated with the northern subcluster, we find a spectral index between 340 MHz and 3 GHz of $\alpha_{340}^{3000}$ = -1.27 $\pm$ 0.05. For the total emission associated with the southern subcluster, we find a spectral index between 340 MHz and 3 GHz of $\alpha_{340}^{3000}$ = -1.25 $\pm$ 0.06. At both frequencies, the integrated radio flux associated with the northern subcluster is comparable within errors to that associated with the southern subcluster. 

\subsubsection{Spectral Index Maps} \label{subsubsec:spixmaps}

\begin{figure*}[ht!]
\centering
\begin{minipage}{0.45\textwidth}
    \includegraphics[width=\linewidth]{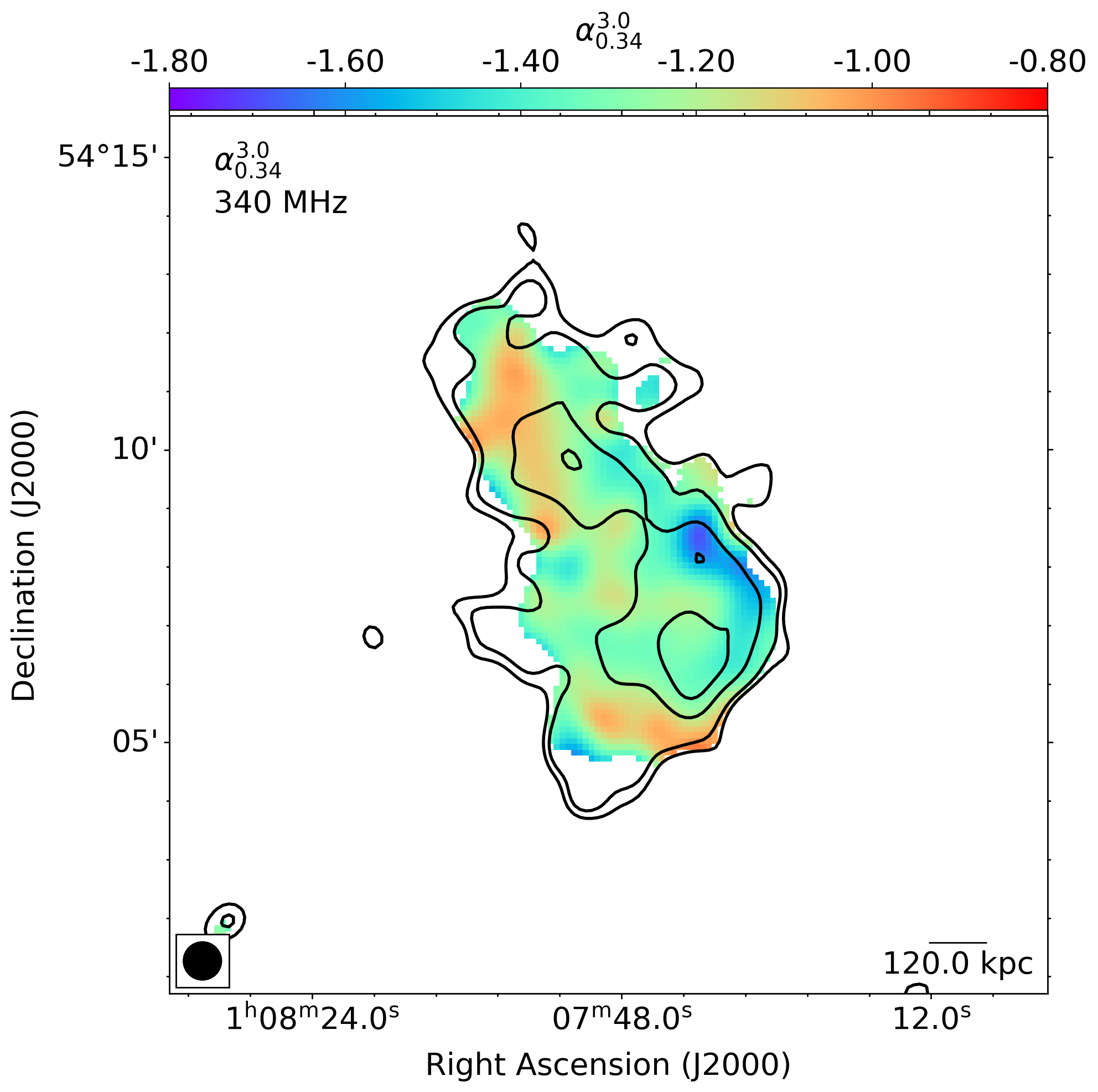}
\end{minipage}
\begin{minipage}{0.45\textwidth}
    \includegraphics[width=\linewidth]{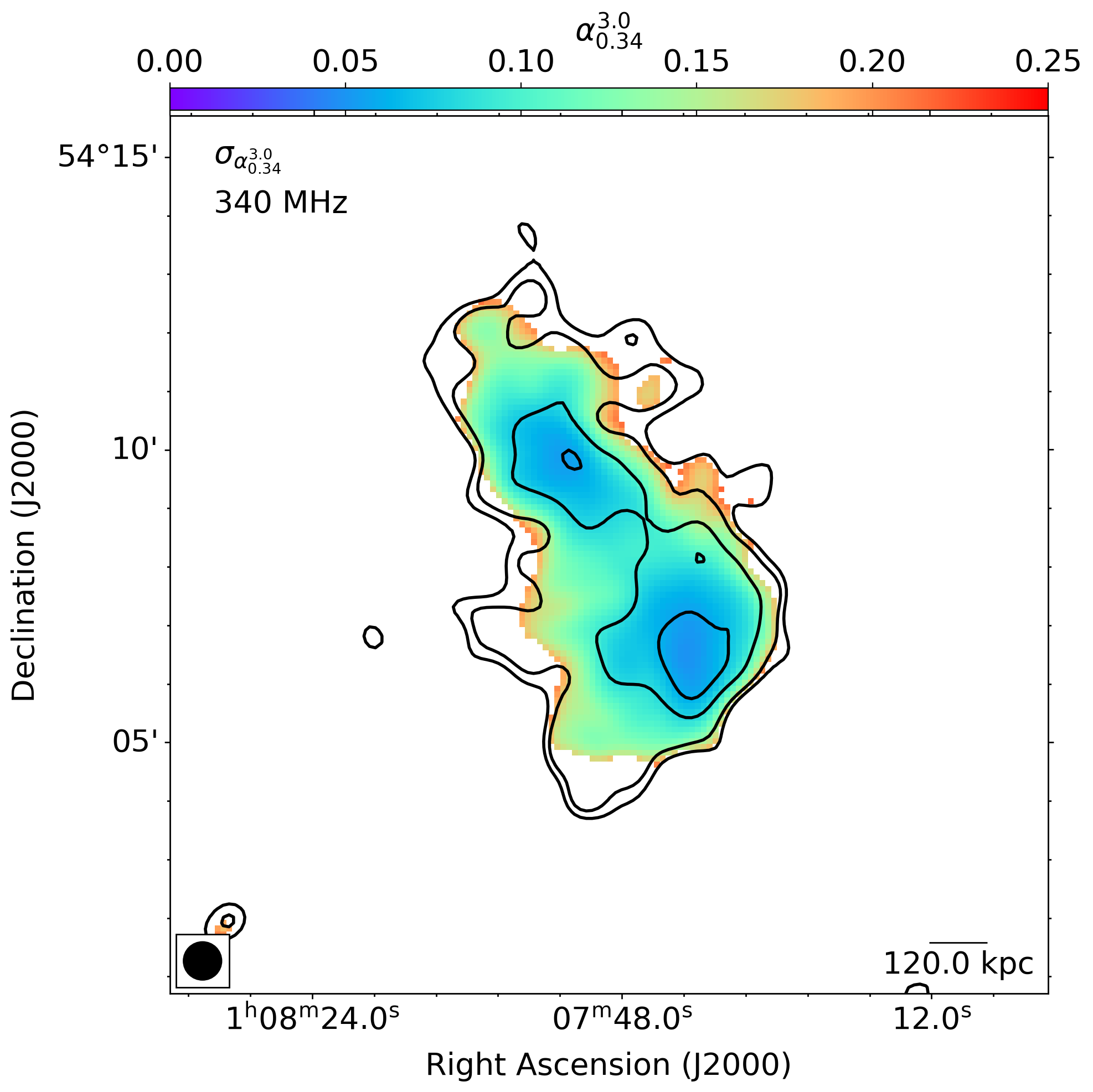}
\end{minipage}
\caption{A (left): Spectral index map between 340 MHz and 3 GHz. Contours from 340 MHz observations at B configuration convolved to a 40'' beam are shown, beginning at 3$\sigma$ and proceeding in integer multiples of $\sqrt{2}$. B (right): Spectral index error map, shown with the same contours from the 340 MHz observations.}
\label{fig:40_SP_spix}
\end{figure*}

We present a spectral index map between 340 MHz and 3 GHz in Figure \ref{fig:40_SP_spix}, generated with \textsc{AIPS} \citep[][]{aipsgreisen}. Each image was primary beam corrected, and the geometry of the lower frequency image was interpolated to the geometry of the higher frequency image. The uv-ranges were matched to ensure that both frequencies were sensitive to emission on the same range of spatial scales. All images were convolved to a common circular beam to correct for differences in beam size. Finally, the spectral index was only calculated for pixels where the flux was above 3$\sigma$ in both images. The resultant spectral index map is presented in Figure \ref{fig:40_SP_spix}A, where it is shown with the 340 MHz contours, again illustrating the sharp drop-off in radio emission to the southwest. The associated spectral index error map is shown in Figure~\ref{fig:40_SP_spix}B. We note that a number of point sources had to be subtracted across the entire cluster, but especially from the northeastern subcluster, and possible source under-subtraction could impact the spectral structure.

Overall, the spectral index map is fairly uniform across the cluster, with an average spectral index of $\alpha \sim -1.3$. We note, however, a few individual features of interest. There are two regions with some slight flattening to the north and the south, but both are consistent with the average cluster value within the errors. The most noticeable feature is steepening coincident with the northwestern USS emission, which is consistent with increased radio emission at lower frequencies. 

\begin{figure}[ht!]
    \centering
     \includegraphics[width=8cm]{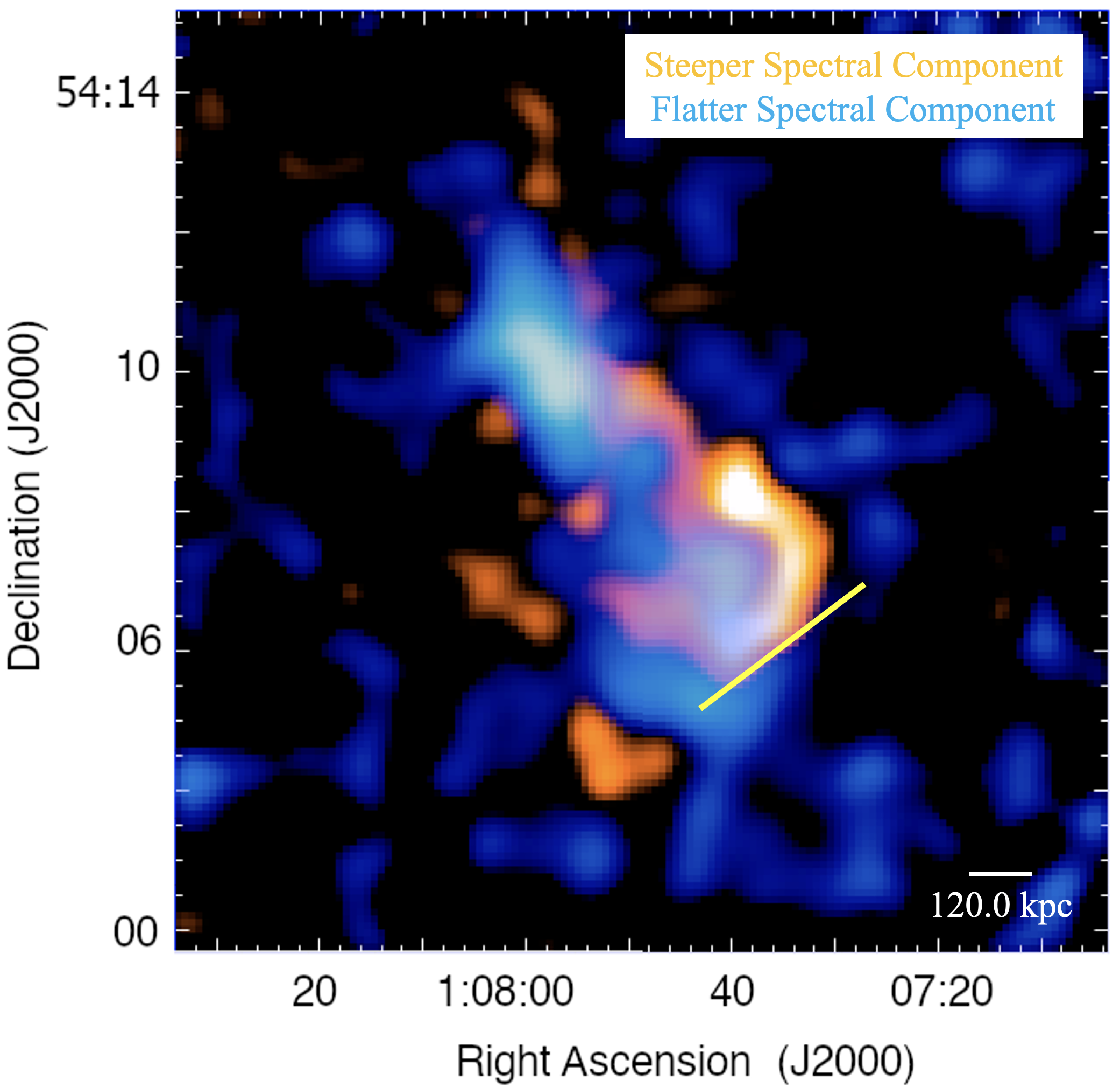}
        \vspace*{1mm}
          \caption{Spectral separation image produced using a technique related to that of spectral tomography in order to highlight the flatter and steeper components in CIZA0107. The flatter emission is represented in blue, while the steeper emission is represented in orange. The yellow line marks the sharp `radio edge' observed at 340 MHz. The images used to generate this map were the matched resolution, point source-subtracted observations, convolved to a 40'' beam. The scale bar represents 120 kiloparsecs.}
    \label{fig:tomography}
\end{figure}

To further investigate the spectral distribution and differentiate the multiple components within CIZA0107, we have produced an image using a technique related to that of spectral tomography \citep[][]{katzstone1993,katzstone1997}, presented in Figure \ref{fig:tomography}. Our goal is to spatially identify and separate the flatter and steeper spectrum components of the emission in CIZA0107. To do this, we generate separate `images' of the two spectral components from the same maps used to generate the spectral index map, though no cut is applied to the signal to noise. Instead, a `flatter spectral component' image is created by subtracting the 340 MHz emission, weighted by a coefficient A, from the 3 GHz emission. Similarly, a `steeper spectral component' image is created by subtracting the 3 GHz emission, weighted by a different coefficient B, from the 340 MHz emission. The coefficients A and B were identified to represent the maximum fraction of each map that could be subtracted without creating negative residuals. The results of this technique are shown in Figure \ref{fig:tomography}, including a yellow line that marks the sharp radio edge observed at 340 MHz. This figure, produced for qualitative purposes only, clearly separates both the NW and SE USS components in orange, in addition to the flatter spectrum diffuse emission in blue. The sharp radio edge seems to trace the boundary of the steeper spectrum emission.

\subsubsection{Ultra-steep Spectrum Regions} \label{subsubsec:USS}

We measured the integrated flux density of each USS regions at 340 MHz on the 40"-resolution image using the 3$\sigma$ 74 MHz contours as a boundary. The values are reported in Table \ref{tab:ussclusterprops}. The errors take into account flux scale uncertainties of 10\% at 340 MHz \citep[][]{NRAO_FluxDensityScale}, in addition to image noise. 

\begin{figure}[ht!]
    \centering
     \includegraphics[width=8cm]{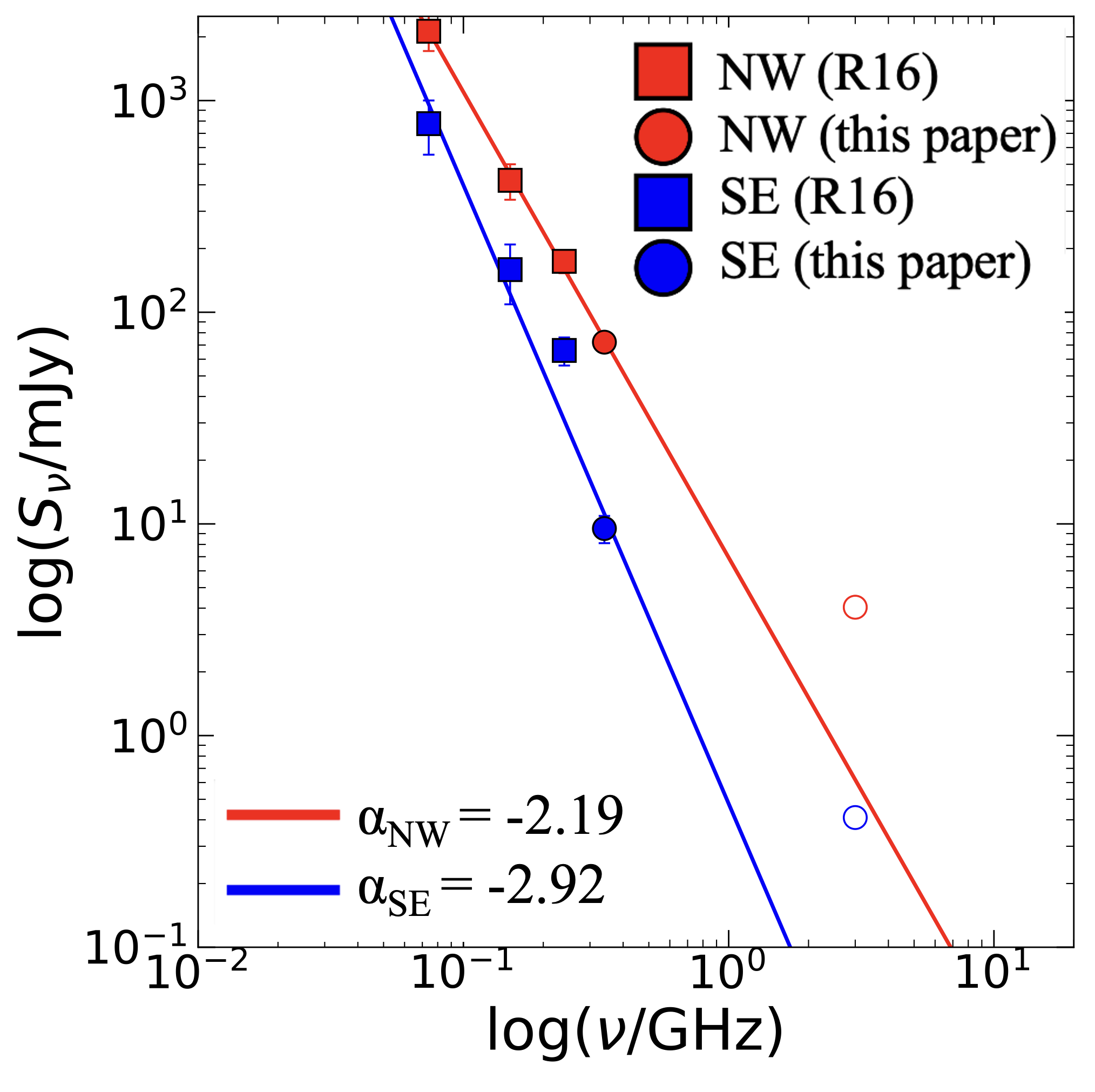}
        \vspace*{1mm}
          \caption{Broadband radio spectra for both the northwestern and southeastern ultra-steep spectrum components, with the northwestern in red and the southeastern in blue. Integrated flux densities from both components are measured at 340 MHz and 3 GHz within the 3$\sigma$ 74 MHz contours. Values from this paper are represented by circles, while values from R16 are represented by squares. Each component has been fit with a standard power law (solid lines). The 3 GHz values (open circles) are not included in the fit as discussed in the text. Resultant spectral indices are listed in the lower left hand corner.}
    \label{fig:uss_flux_figs}
\end{figure}

Figure~\ref{fig:uss_flux_figs} displays our measurements (filled circular points) and data presented in R16 from the VLSSr survey at 74 MHz and the Giant Metrewave Radio Telescope at 150 MHz and 240 MHz, all represented with square points. Least squares fitting was performed for each component (NW in red, SE in blue) following the approach described in \cite{patil2022}, making use of the \texttt{Radio Spectral Fitting}\footnote{https://github.com/paloween/Radio\_Spectral\_Fitting} tools, in which the fit is weighted by the error associated with each point \cite[][]{patil2021}. The standard power law fit results in spectral indices of $\alpha$ = -2.19 $\pm$ 0.07 for the northwestern component, and $\alpha$ = -2.92 $\pm$ 0.08 for the southeastern component (Tab.~\ref{tab:ussclusterprops}). These spectral indices are consistent with those reported in R16 within $2\sigma$. 

As noted in Section~\ref{subsec:radiocontinuum}, the USS regions are not clearly detected at 3 GHz. In Figure~\ref{fig:uss_flux_figs} we plot as empty circles the 3 GHz flux density measured within the region occupied by the 74 MHz 3$\sigma$ USS contour
($S_{3 GHz}$ = 4.04 $\pm$ 0.22 mJy and 0.41 $\pm$ 0.05 mJy for the NW and SE USS regions, respectively). It is clear that these values are well above the best-fit power law, suggesting that the 3GHz emission is dominated by the flatter-spectrum diffuse emission in the SW subcluster. This is also supported by the spectral tomography shown in Figure \ref{fig:tomography} and indicates that there are two superimposed components: the diffuse emission associated with the SW subcluster, which dominates in the GHz regime, and the emission from the USS regions, which dominates between 74 and 340 MHz.

\begin{deluxetable}{ccccc}
\tablenum{4}
\tablecaption{Ultra-steep Spectrum Regions Properties}
\tablewidth{0pt}
\tablehead{
\colhead{} & \colhead{$S_{340 MHz}$} & \colhead{$\alpha_{74 MHz}^{340 MHz}$} & \colhead{$\theta(r_p)_{74 MHz}$}\\
\colhead{} & \colhead{mJy} & \colhead{} & \colhead{''(kpc)}}
\startdata
NW & 72.2$\pm$7.4 & -2.19$\pm$0.07 & 260(500)\\
SE & 9.5$\pm$1.4 & -2.92$\pm$0.08 & 140(265)\\
\enddata
\caption{\textbf{Notes}: Column 1 - Region. Column 2 - Integrated flux $\pm$ error in mJy at 340 MHz. Column 3 - Best-fit spectral index between 74 MHz and 340 MHz from Figure \ref{fig:uss_flux_figs}. Column 4 - Angular and projected physical sizes of each USS region, measured at 74 MHz within the 3$\sigma$ contours, along the axis of the merger.}
\label{tab:ussclusterprops}
\end{deluxetable}

\section{X-ray Edge Analysis} \label{sec:xray}

\begin{figure*}[ht!]
\centering
\begin{minipage}{0.45\textwidth}
    \includegraphics[width=\linewidth]{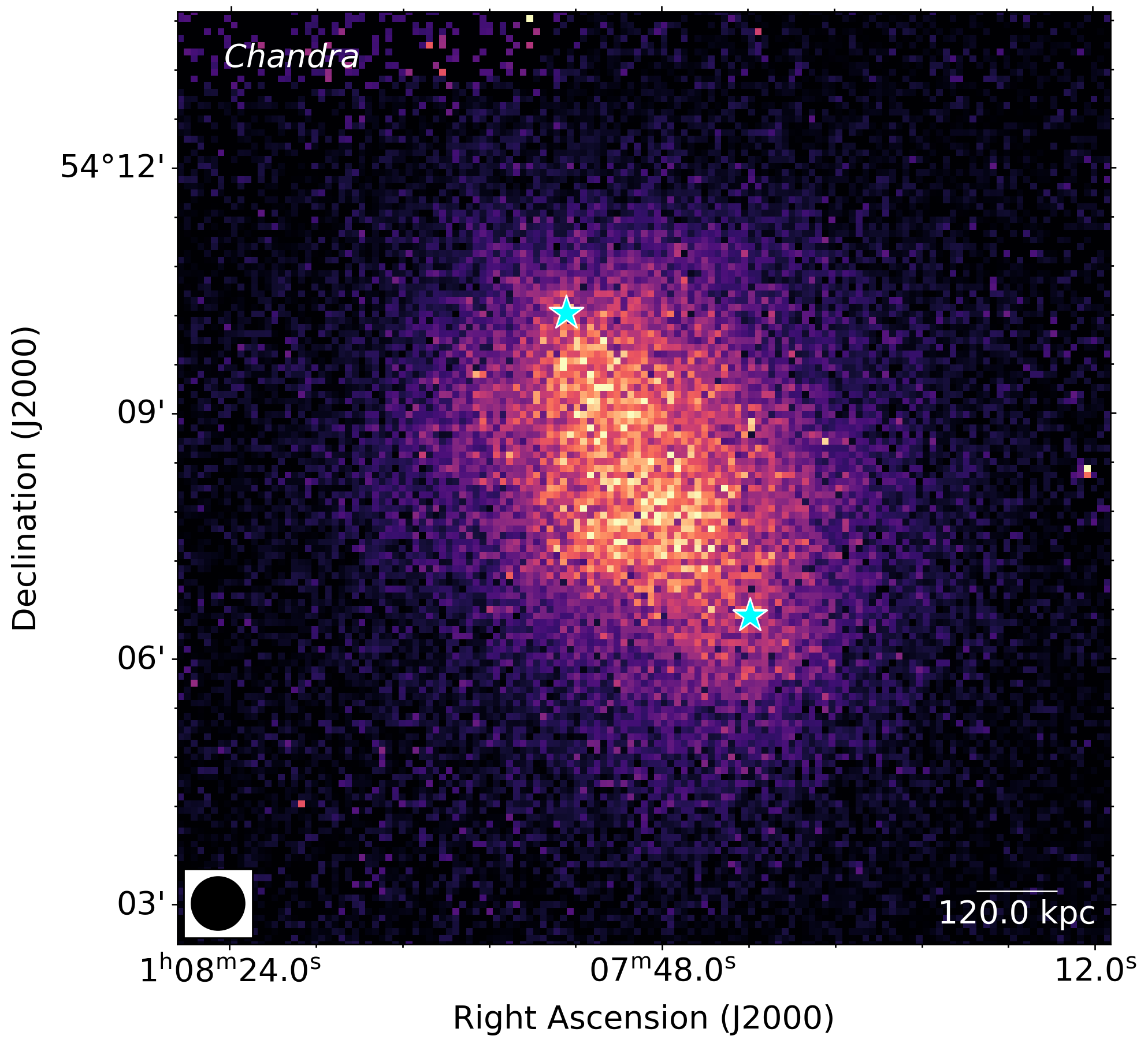}
\end{minipage}
\begin{minipage}{0.45\textwidth}
    \includegraphics[width=\linewidth]{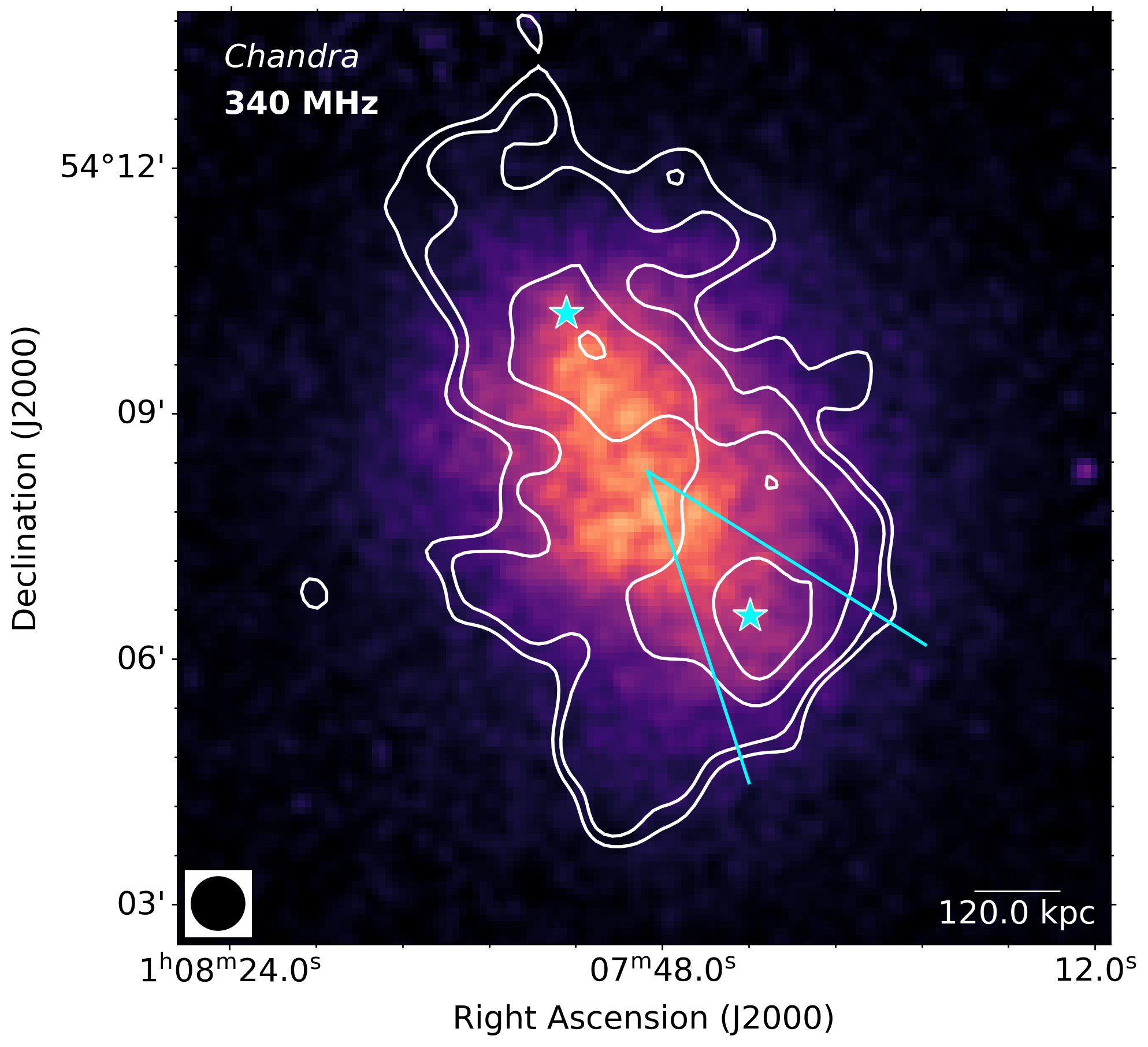}
\end{minipage}
\begin{minipage}{0.45\textwidth}
    \includegraphics[width=\linewidth]{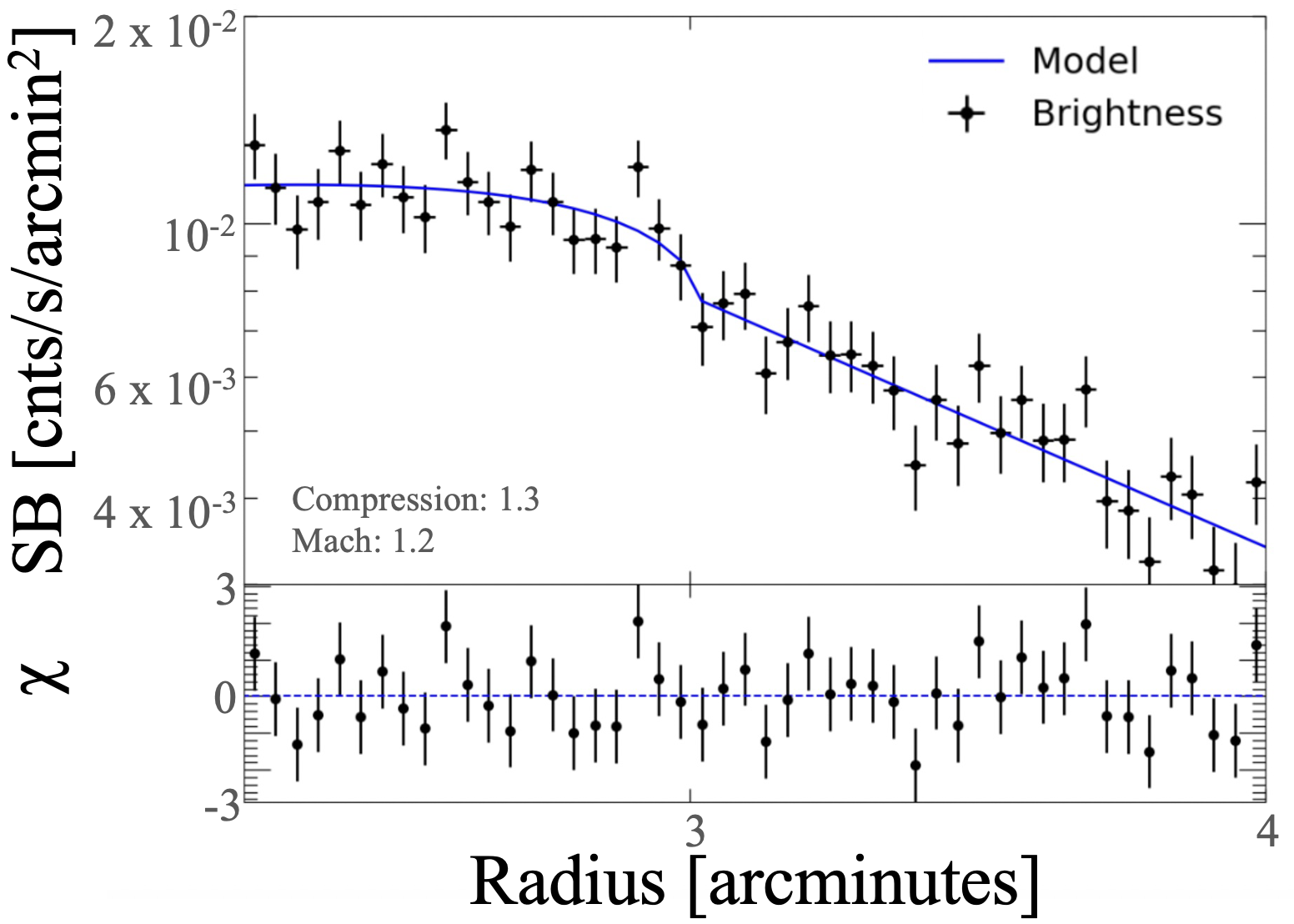}
\end{minipage}
\begin{minipage}{0.45\textwidth}
    \includegraphics[width=\linewidth]{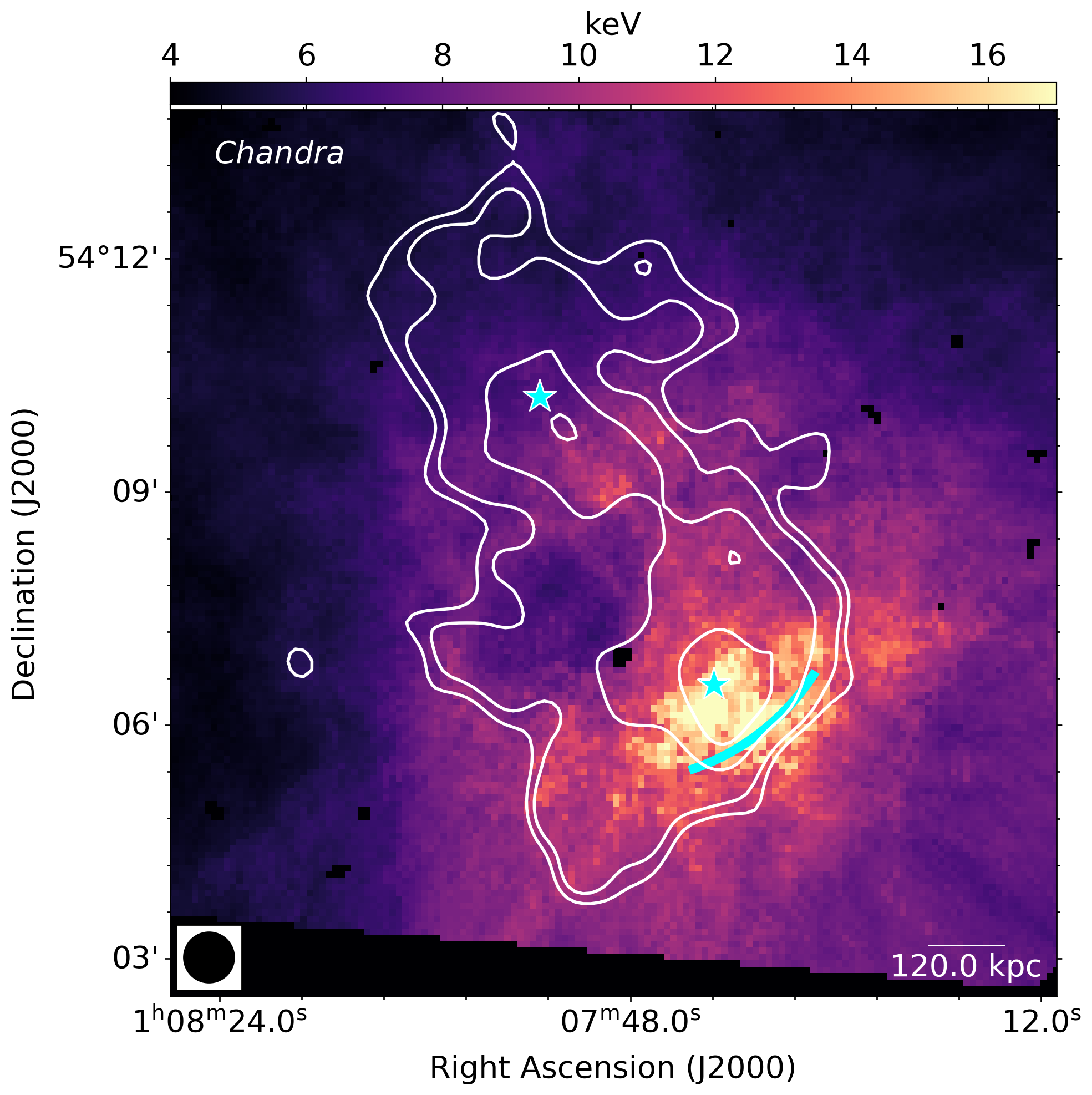}
\end{minipage}
\caption{A (top left): 160 ks, point-source-subtracted, exposure-corrected and background-subtracted  \textit{Chandra} X-ray image in the 0.5-7 keV band, binned by 10 pixels to a new pixel size of 4.92''. BCGs are marked with cyan stars. B (top right): the bin-10 pixel \textit{Chandra} image, now smoothed with a circular Gaussian with a kernel size of 1$\sigma$ = 1 pixel = 4.92''. The white contours trace the 340 MHz B-configuration VLA emission, convolved to a 40'' beam, beginning at 3$\sigma$ and proceeding in multiples of $\sqrt{2}$. The wedge used for the X-ray surface brightness profile extraction is marked in cyan. C (bottom left): 0.5-7 keV \textit{Chandra} X-ray surface brightness profile extracted in the wedge shown in B, using logarithmically sized bins with a minimum size of 1''. The profile was modeled with a broken power-law projected on the plane of the sky. The model identifies a discontinuity at 3'. D (bottom right): \textit{Chandra} temperature map presented in R16 in units of keV, overlaid with 340 MHz B-configuration VLA contours matching those in panel B. The BCGs are marked with cyan stars. The physical scale is 1.95 kiloparsecs per arcsecond.}
\label{fig:chandra_radio}
\end{figure*}

CIZA0107 was observed 5 times by \textit{Chandra} ACIS-I for a total of 160 ks (ObsIDs 15152, 21557, 22879, 23065, 23210; R16, Randall et al. in preparation). A point source-subtracted, exposure-corrected, 0.5-7.0 keV \textit{Chandra} image from the combined \textit{Chandra} observations was presented in \cite{Finner_2023}. In Figure \ref{fig:chandra_radio}A, we show the \textit{Chandra} image of CIZA0107 in the 0.5-7.0 keV band, binned by 10 pixels to a new pixel size of 4.9'', with the BCGs marked with blue stars. We defer to Randall et al. in prep. for details on the \textit{Chandra} data reduction and image preparation.

In Section \ref{subsec:radiocontinuum}, we identified a sharp radio edge in the new high-resolution radio images at 340 MHz. Here, we exploit the superb angular resolution of \textit{Chandra} to search for a possible X-ray surface brightness discontinuity at the position of the radio edge. A spatial coincidence between radio relics/halo-edges and X-ray shock fronts is often observed in the outer regions of a merger \citep[][]{giacintucci2008, botteon2016shock115}. For further discussion of the existence of a radio relic or radio halo-shock edge, see Section \ref{subsec:radioorigin}. 

In the case of CIZA0107, Figure \ref{fig:chandra_radio}B presents the \textit{Chandra} image, binned by 10 pixels and further smoothed with a circular Gaussian with a kernel size of 1$\sigma$ = 4.92'', overlaid with the 340 MHz VLA contours. Guided by the radio observations, we defined a wedge that encompasses the sharp radio edge, marked in cyan in Figure \ref{fig:chandra_radio}B, within which we extracted and modeled the X-ray surface brightness profile using \textsc{PyProffit}\footnote{https://github.com/domeckert/pyproffit} \citep[][]{eckert2020}. 

The profile was extracted in small annuli from the center of the wedge out to a large cluster radius encompassing the radio edge using logarithmically-sized bins with a minimum size of 1'' for each bin. The resulting surface brightness profile is shown in Figure \ref{fig:chandra_radio}C. A weak discontinuity is visible at a radius of $\sim$ 3', and was fit using a broken power law projected onto the plane of the sky, with a reduced $\chi^2 = 1.004$. The model-calculated location of the discontinuity ($r_{\rm model}=3.02'$) is marked with a cyan curve in Figure \ref{fig:chandra_radio}D, and is coincident with the sharp radio edge at 340 MHz (within the 40'' spatial resolution of the radio image). We also tried to fit the profile in the wedge using a single $\beta$-model, which provided a core radius of $r_c$ = 5.6' and $\beta$ = 1.7, with a reduced $\chi^2$ = 1.29. Thus, the broken power law provides a statistically better fit.

Figure \ref{fig:chandra_radio}D presents the \textit{Chandra} gas temperature map from R16, which highlights an increase in temperature coincident with the location of both the sharp radio edge and the X-ray surface brightness discontinuity, as is expected for a shock. Using a wider wedge than we used here, R16 reported a possible density discontinuity in this high-temperature region with a density jump of $\sim$1.2. The broken power law model fit to our profile in Figure \ref{fig:chandra_radio}C gives a similar compression factor (ratio of post-shock to pre-shock density) of $C=1.3\pm0.12$. Using the Rankine-Hugoniot jump conditions \citep[][]{landau1959} to account for the hydrodynamical properties across a bow shock, the following expression is used to determine the Mach number:
\begin{center}
\begin{equation} \label{eq:1}
    \frac{1}{C} = \frac{1}{\mathcal{M}^{2}}\frac{2}{\gamma + 1} + \frac{\gamma - 1}{\gamma + 1},
\end{equation}
\end{center}
where $\frac{1}{C}$ is equivalent to the gas density ratio $\frac{\rho_{2}}{\rho_{1}}$, and $\gamma$ is 5/3. This gives a Mach number $\mathcal{M}=1.2\pm0.6$, corresponding to a weak shock. However, the rapid decline in temperature visible in Figure \ref{fig:chandra_radio}D, in the region of the shock suggests a potentially higher Mach number. Since there are indications of a significant line of sight component to the merger \citep[][]{Finner_2023}, the density jump estimate, and thus the Mach number, can be artificially reduced due to projection effects (see Randall et al. in preparation, for a detailed analysis).

\begin{figure}[ht!]
    \centering
     \includegraphics[width=8cm]{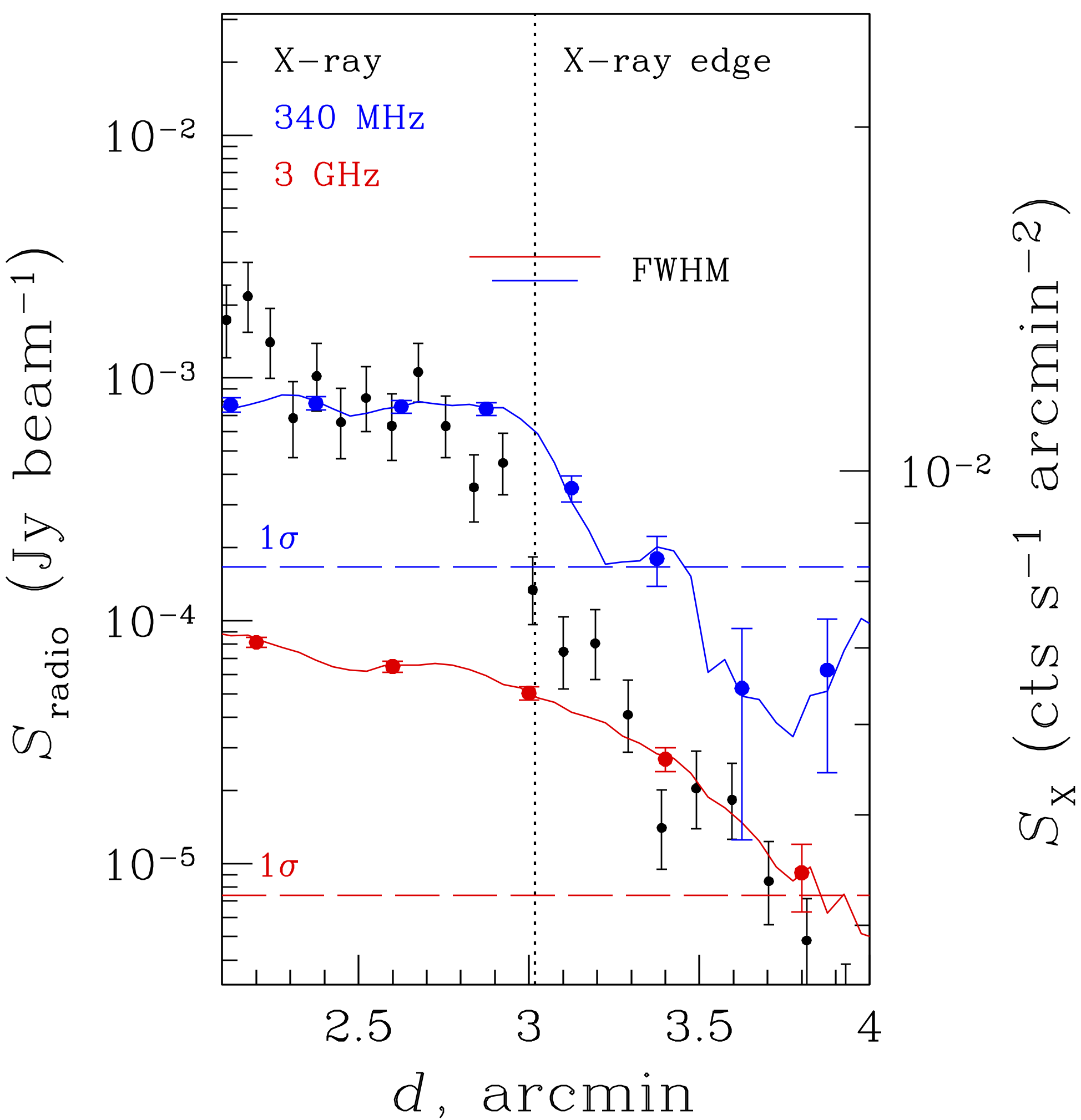}
        \vspace*{-1mm}
          \caption{The radio (blue, red) and X-ray (black) surface brightness profiles. The X-ray values were extracted using logarithmically-spaced bins with a minimum size of 1 pixel = 0.492''. The radio values plotted as points use radial bins as wide as FWHM ( 15'' at 340 MHz and 24'' at 3 GHz). The radio values plotted as lines use radial bins of 1 pixel = 3''. The 1$\sigma$ noise levels in the radio images are indicated at both frequencies. The position of the X-ray surface brightness discontinuity (represented by the cyan curve in Figure \ref{fig:chandra_radio}D) is indicated by the vertical dotted line. The blue and red lines correspond to regions that are not beam-independent, while the points correspond to regions that are beam-independent. 
}
    \label{fig:xrayradioprofile}
\end{figure}

In Figure \ref{fig:xrayradioprofile}, we compare the radio emission profiles at both 340 MHz and 3 GHz across the position of the X-ray discontinuity. Surface brightness profiles were extracted within the same wedge that is presented in Figure \ref{fig:chandra_radio}B. The X-ray values, shown in black points, were extracted using logarithmically-spaced bins with a minimum size of 1 pixel = 0.492''. The blue and red points represent the radio values extracted using radial bins as wide as the full width half maximum (FWHM; 15'' at 340 MHz and 24'' at 3 GHz). The blue and red lines represent the radio values extracted using radial bins of 1 pixel = 3''. The 1$\sigma$ noise levels in both radio maps are indicated with horizontal dashed lines. The position of the X-ray surface brightness discontinuity is shown by a vertical dotted black line, and is aligned with the location of the cyan curve in Figure \ref{fig:chandra_radio}D. The 340 MHz profile shows a clear edge coincident with the X-ray edge. However, the 3 GHz radio profile does not show any corresponding edge-like feature at that location and in fact extends beyond the drop-off seen at 340 MHz. We confirm that a sharper drop off at 340 MHz compared to at 3 GHz is maintained when the profile is extracted at the lower spatial resolution of the 3 GHz image. This edge feature is further discussed in Section \ref{subsubsec:southwest}. 

\section{Discussion} \label{sec:discussion}

\subsection{Diffuse Radio Emission} \label{subsec:radioorigin}

The new VLA observations confirm the complex dynamical state of CIZA0107. They show a dramatically disturbed merger system with a merger axis along the NE-SW direction. At both 340 MHz and 3 GHz, there is diffuse emission associated with the two subclusters, spatially consistent with the X-ray surface brightness and optical density peaks. It is possible that the diffuse emission is tracing two radio halos, two radio relics seen projected onto the central regions of the cluster, or some combination thereof. Double radio relics are not uncommon; for example, Abell 1240 \citep[][]{hoang2018} or Abell 3667 \citep[][]{degasperin2022}. Double radio halos are instead extremely rare, with only two cases reported so far; i.e., Abell 399-401 \citep[][]{murgia2010} or Abell 1758N-1758S \citep[][]{botteon2018}. Both are widely-separated, pre-merging cluster pairs, unlike CIZA0107, which is a post-merger system. In the following sections, we discuss the diffuse radio emission features in CIZA0107.

\subsubsection{Southwestern Subcluster, Radio Edge, \& USS Regions} \label{subsubsec:southwest}

The southwestern subcluster hosts diffuse radio emission on scales of hundreds of kpc, with a steep radio spectral index ($\alpha \sim$ -1.25). The radio emission is coincident with the X-ray emission, suggesting that it could be a radio halo associated with this subcluster. Our new radio images at 340 MHz reveal that this diffuse component displays a sharp edge at low frequency that is not seen at higher frequencies, illustrated in Figure \ref{fig:xrayradioprofile}. At 340 MHz, the sudden drop-off is coincident with both a discontinuity in the X-ray surface brightness and a region of increased gas temperature, suggesting a shock front. The Mach number derived from the X-ray surface brightness discontinuity is $\sim$ 1.2. However, it is likely that there is a significant line of sight component to the merger \citep[][]{Finner_2023}, and thus the shock is seen partially face on, weakening the measured value of the Mach number.

An edge in the diffuse radio emission coincident with an X-ray shock front may indicate a radio relic or radio halo-shock edge, similar to that detected in the Coma Cluster \citep[][]{brown2011}, the Bullet Cluster \citep[][]{shimwell2014}, or Abell 520 \citep[][]{wang2018}. However, in CIZA0107, the edge is observed only at 340 MHz, whereas the 3 GHz diffuse emission does not show any corresponding edge-like feature at that location and in fact it extends beyond the shock front. This behavior is in tension with both a relic or radio halo-shock edge interpretation. For this reason, we propose an alternative scenario in which the edge structure at 340 MHz is an extension of the USS region to the northwest, as also suggested by the spectral tomography in Figure \ref{fig:tomography}.

The observed USS emission is consistent with a fossil, non-thermal electron population (for instance from aged lobes of active galactic nuclei) that has been adiabatically compressed and re-energized by the passage of the shock, as is suggested for radio phoenixes \citep[][]{kempner2004}. Due to its steep spectrum, this emission is undetected at 3 GHz, where flatter spectrum emission associated with a possible radio halo dominates. This scenario might also explain why the USS regions are smaller than and misaligned from the larger, flatter diffuse components, as the fossil electron populations dominating the USS regions are distinct from those dominating the more extended diffuse components. Though we observe no such radio galaxy that might exist as a source of fossil electrons for either the observed sharp radio edge or the USS regions, such radio lobes would have had significant time to detach and fade below observational limits, as is consistent with other such systems \citep[][]{murgia2005,dutta2023}.

\subsubsection{Northeastern Subcluster} \label{subsubsec:northeast}

The emission associated with the northern subcluster of CIZA0107 has fewer distinguishing features available to help categorize the emission. Unlike the emission associated with the southern subcluster, there is no indication in the radio observations of a radio edge associated with the northern subcluster. Though the northeastern emission could be a radio halo, given the line of sight component to the merger, we cannot rule out that the diffuse emission is a relic that is being seen partially face on \citep[][]{wonki2024}. We note the lack of spectral features in the spectral index map across the northeastern component. This is somewhat surprising; if the observed diffuse emission is a turbulence-driven radio halo, in principle, the inhomogeneity of turbulent re-acceleration would be expected to cause spectral variation \citep[e.g., Abell 754;][]{botteon2024}. However, spectral index maps have been found to be relatively smooth in, for example, the massive merging clusters Abell 2744 \citep[][]{pearce2017} and the Toothbrush \citep[][]{rajpurohit2018}. 

In the case of CIZA0107, this could imply that the turbulent energy flow and turbulent damping on the 40'' ($\sim 80$ kiloparsecs) beam-size scales of the images is essentially constant when integrated along the line of sight. Thus, the turbulent energy does not increase or decrease significantly across the overall cluster merger. While detailed simulations are required to draw meaningful conclusions from this, it is possible that these constraints could be important to models of turbulence and re-acceleration in the ICM. However, the apparent smoothness of the spectrum may also be a result of the sensitivity and resolution limitations of the observations.

\subsubsection{A Radio Halo Pair?} \label{subsubsec:halopair}

Here we consider the possibility that each of the two subclusters in CIZA0107  hosts a radio halo and compare their radio/X-ray properties to known scaling relations and correlations for radio halos.

A correlation has been found between the power of radio halos and the total mass of their host clusters \citep[][]{cassano2013}. The integrated flux density of the radio emission associated with CIZA0107 (Table \ref{tab:subclusterprops}) was used to calculate the total radio luminosity at 1.4 GHz assuming a spectral index of $-1.3$ (Table \ref{tab:subclusterprops}): log($L_{1.4GHz}$) = 24.10 W/Hz$^{-1}$. Using this luminosity and the total cluster mass $M_{500}$\footnote{Total mass within $R_{500}$, where $R_{500}$ is the radius within which the cluster mean total density is 500 times the critical density at the cluster redshift.} from \cite{Finner_2023}, we explore the location of CIZA0107 on the $P_{1.4 GHz}$-$M_{500}$ presented in Figure 4 in \cite{cuciti2021}. The cluster was found to be slightly under-luminous in the radio compared to the best-fit relation shown in \cite{cuciti2021}, but its position in the plot is still in agreement with the location of few radio-halo clusters with a similar total mass. 

Radio-halo clusters often display a point-to-point positive correlation between the X-ray and radio surface brightness, indicating a spatial link between the distribution of both emissions \citep[][]{govoni2001,govoni2001b,bonafede2022,rajpurohit2023}. The slope of this correlation is typically found to be sub-linear (see \cite{balboni2024} and references therein). 

\begin{figure*}[ht!]
    \centering
     \includegraphics[width=18cm]{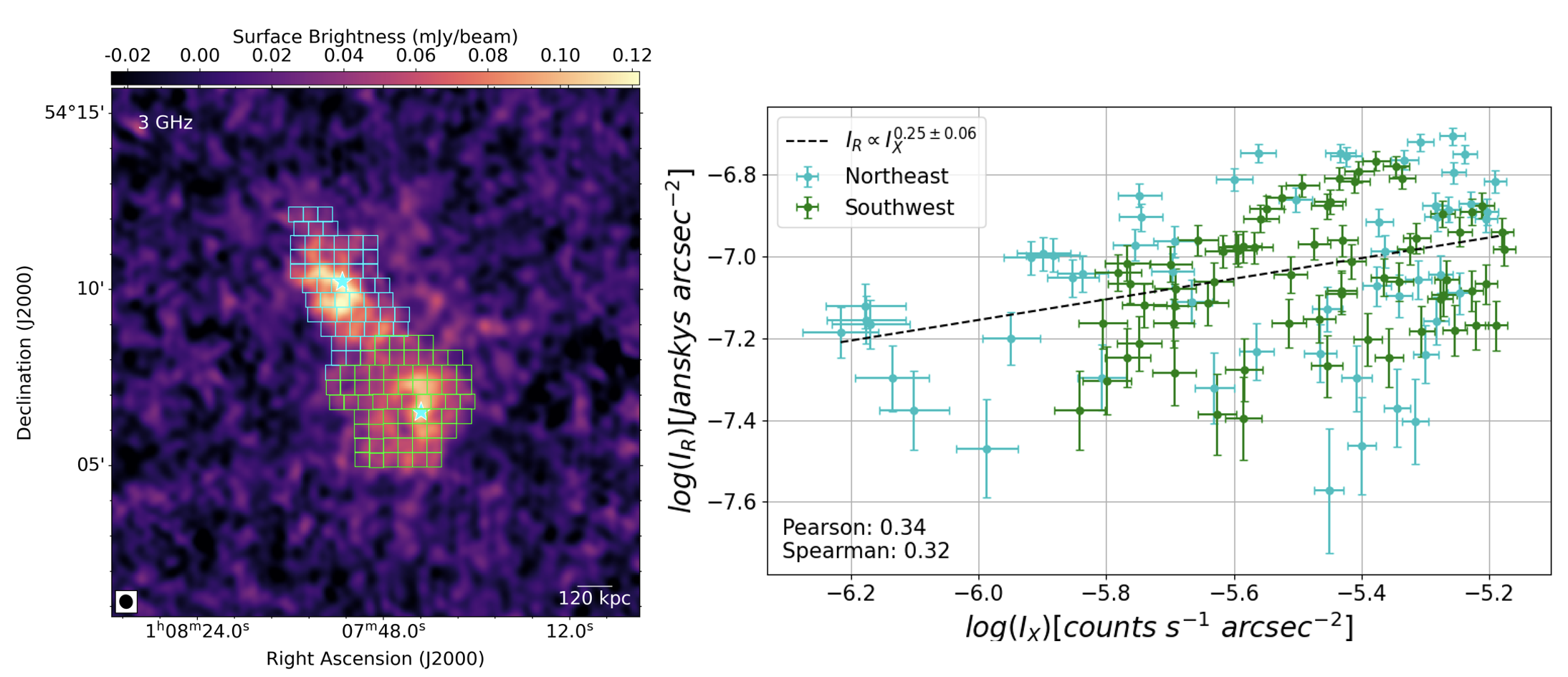}
        \vspace*{1mm}
          \caption{A (left): The 3 GHz point source-subtracted radio map, shown with the cells used for the point-to-point analysis. All cells were drawn within the 3$\sigma$ contour and are beam-independent (25''$\times$25'' size). The BCGs are marked with cyan stars. B (right): $I_{R}-I_{X}$ relation for the diffuse emission associated with the northern and southern subclusters in CIZA0107. The surface brightness was extracted from the 3 GHz native resolution point source-subtracted image. The northern emission is marked in blue, the southern emission in green. The \textsc{Linmix} best-fit relation is shown with a dashed black line, and the best fit is listed in the legend, along with the Pearson and Spearman correlation coefficients.}
    \label{fig:ptp}
\end{figure*}

Using the X-ray and radio data presented in this paper, we check whether such a spatial correlation exists for the diffuse emission associated with the subclusters in CIZA0107. We use a background-subtracted and exposure-corrected \textit{Chandra} image in the 0.5-7.0 keV energy band with point sources subtracted out. For the radio, we use the native resolution, point-source subtracted radio map at 3 GHz (Figure \ref{fig:ciza_subims}B) to limit the contamination from the superposed USS emission, which becomes much brighter at 340 MHz. A grid was drawn within the 3$\sigma$ contour of the radio map using beam-independent 25''$\times$25'' cells (Figure \ref{fig:ptp}A), and the average radio ($I_{R}$) and X-ray ($I_{X}$) surface brightness values were measured in each cell. 

Figure \ref{fig:ptp}B presents the result of the point-to-point analysis. Blue points and cells correspond to the northern emission, while green to the southern. To check the existence of a possible positive correlation, the data were fit with a power law $I_{R} \propto I_{X}^{b}$, where \textit{b} is the slope of the correlation. The \textsc{Linmix} package \citep[][]{kelly2007} was used for the fitting, and both the Spearman and Pearson coefficients were computed to assess the strength of the correlation. 

We find a moderate X-ray/radio surface brightness correlation for the whole diffuse emission in CIZA0107. The correlation slope \textit{b} is found to be 0.25, with Pearson and Spearman correlation coefficients of 0.34 and 0.32 respectively (also listed in Fig.~\ref{fig:ptp}B). The tests also return p-values, the two-sided significance of the rank correlation coefficient, under the null hypothesis of no correlation between the X-ray and the radio surface brightness. For the Spearman correlation coefficient, we find p = 0.0003, and for the Pearson correlation coefficient, we find p = 0.0002. These small p-values indicate a statistically significant positive correlation. For the individual subclusters, we find correlation slopes of $\sim$0.25, with Pearson and Spearman coefficients of 0.34 and 0.40 respectively, and similar p-values, indicating a moderate correlation at a high level of statistical significance. 
 
In general, radio-halo clusters exhibit a relatively strong sub-linear correlation with a slope steeper than the one we measure for CIZA0107 and with higher Spearman correlation coefficients ($\geq 0.5$; e.g., \cite{balboni2024} and references therein). However, weaker correlations with slopes as flat as in CIZA0107 have been reported for a few clusters \citep[e.g.,][]{campitiello2024}. In CIZA0107, it is possible that the presence of further (non-halo) components of diffuse emission along the line of sight may weaken an existing stronger point-to-point correlation and affect its slope (see e.g. \cite{balboni2024}, who found that correlations appear stronger when excluding projected contaminant regions from the images of the radio halos). 

The position of CIZA0107 in the $P_{1.4 GHz}$-$M_{500}$ is in agreement with the observed scaling relation for radio halos, thus suggesting that the observed diffuse emission may be a pair of radio halos. The finding of a moderate, but statistically significant, X-ray to radio point-to-point correlation may support this interpretation. However, while it is possible that CIZA0107 contains a pair of radio halos, we cannot rule out a scenario in which the diffuse emission arises from two radio relics projected onto the central regions of the cluster, as was suggested in R16. 

\subsubsection{A Projected Double Relic?} \label{subsubsec:relicpair}

In a double-relic system scenario, we can use the radio luminosity of the two individual subclusters and the total cluster mass to check the location of CIZA0107 on the $P_{1.4 GHz}-M_{500}$ plot for double relics presented in Figure 7A of \cite{degasperin2014}. For each relic pair, this relation compares the radio luminosity of each relic to the total mass of the host cluster. The study found that for single and double relic systems there was a strong correlation between these two measures. In the case of CIZA0107, if the observed diffuse emission is associated with two projected relics, we find that both relics would be about an order of magnitude under-luminous in the radio with respect to the best-fit correlation.

We note, however, that the relation found by \cite{degasperin2014} is not based on a statistical sample, but rather on an heterogeneous collection of clusters with double radio relics. A revision of this relation was presented by \cite{jones2023} (their Fig.~6) using a mass-selected cluster sample from LOFAR LOTSS DR2 at 150 MHz and both single and double relics. With respect to revised relation, the putative double relic in CIZA0107 remains a significantly under-luminous outlier. However, the individual merger dynamics, including impact parameter, mass ratio, ICM magnetic field configuration, and projection effects may impact the strength and (re)acceleration efficiency of injected shocks, and thus lead to both over and under-luminous relics with respect to the radio luminosity-mass relation \citep[][]{pal2024}. 

\section{Conclusions \& Future Work} \label{sec:futuresummary}

We present new VLA radio observations at two frequencies of the dissociative, binary, post-core passage galaxy cluster merger CIZA J0107.7+5408. Using the new radio observations, in addition to existing radio and X-ray observations from \cite{Randall_2016}, \cite{Finner_2023}, we find:

\begin{itemize}
    \item New radio observations at 340 MHz and 3 GHz confirm diffuse radio emission on a scale of $\sim$0.5 Mpc in each of the NE and SW subclusters. This emission encompasses the BCGs in both subclusters. At 340 MHz, we also detect emission from two ultra-steep spectrum regions initially identified by R16 at 74 MHz. The USS regions are located to the northwest and southeast of the diffuse radio emission peak in the SW subcluster.
 
    \item At both 340 MHz and 3 GHz, the integrated radio flux associated with the NE subcluster is comparable (within errors) to that associated with the SW subcluster (flux density ratio of $\sim$1:1). Consequently, the NW and SE subclusters have a similar spectral index ($\alpha\sim -1.3$). For the USS regions, we measure steeper spectral slopes between 74 MHz and 340 MHz of $\sim -2.2$ and $\sim -2.9$ for the NW and SE regions, respectively. Spectral tomography clearly separates both the USS components from the flatter spectrum diffuse emission, thus suggesting that these components are spatially overlapping. 

    \item A spectral index map generated between 340 MHz and 3 GHz after removal of compact sources reveals the diffuse emission has a approximately uniform spectrum with an average $\alpha \sim -1.3$. A region with steeper emission is detected at the northwestern USS emission.
    
    \item The diffuse emission associated with the SW subcluster shows a sharp radio edge at 340 MHz. Using \textit{Chandra} observations, we have identified an X-ray surface brightness discontinuity at the radio edge and spatially coincident with a high-temperature region reported by R16. Fitting the X-ray discontinuity with a broken power law projected onto the plane of the sky, we obtain a Mach number of $\sim 1.2$. Since there are indications of a significant line of sight component to the merger, the Mach number estimate can be artificially reduced due to projection effects.

    \item Binary mergers are expected to launch a pair of shocks in opposite directions and thus we may expect a counter shock to be located in the NE region of the cluster. We see no clear evidence of the counter shock in the X-ray surface brightness or temperature maps. Further results from a detailed X-ray analysis will be presented in Randall et al. (in preparation).
        
    \item The 3 GHz diffuse emission does not show any feature at the location of the 340 MHz radio edge and X-ray shock front, and in fact it extends beyond the shock. The spectral tomography suggests that the 340 MHz edge is is an extension of the USS region to the northwest. The observed USS emission may arise from a fossil, non-thermal electron population (for instance from aged AGN lobes) that has been adiabatically compressed and re-energized by the passage of the shock.

    \item The larger-scale diffuse emission in CIZA0107 may be tracing two radio halos or two projected radio relics. For the case of two halos, we find that the cluster total emission is in agreement with the $P_{1.4 GHz}-M_{500}$ scaling relation for radio halos. We also find a moderate but statistically significant $I_{R}-I_{X}$ point-to-point correlation. The slope of 0.25 is flatter than is usually found for radio halos, possibly indicating the presence of further (non-halo) diffuse components along the line of sight that may weaken a stronger point-to-point correlation and affect its slope. If, instead, the cluster hosts two projected relics, we find them to be under-luminous by about an order of magnitude in comparison to the $P_{1.4 GHz}-M_{500}$ best-fit correlation for double-relic systems.
     
\end{itemize}

There is considerable future work to be done with CIZA0107. Higher resolution and sensitivity radio observations, including polarimetry, could be used to better separate the overlapping diffuse components both spatially and spectrally. Higher frequency radio observations may help to constrain the nature (relic/halo/phoenix) of the observed emission and identify the physical acceleration mechanism at work. Numerical simulations may help to understand the dynamics and merger geometry of CIZA0107. Finally, as a dissociative, binary cluster merger with two nearly equal mass subclusters, CIZA0107 has the potential to be used to place constraints on dark matter. 

\begin{acknowledgements}
We thank the anonymous referee for helpful suggestions that have improved the paper. E. S. gratefully acknowledges support from George Mason University's Doctoral Research Scholars program. SWR acknowledges support from the Smithsonian Institution, the Chandra X-ray Center through NASA contract NAS8-03060, and Chandra grant GO9-20118X. This research made use of Astropy, a community-developed core Python package for Astronomy (\cite{astropy}), $\mathrm{TOPCAT}$ (\cite{topcat}), the Common Astronomy Software Application (\cite{casanew2022}), and the Python Blob Detector and Source Finder (\cite{pybdsf}). The National Radio Astronomy Observatory is a facility of the National Science Foundation operated under cooperative agreement by Associated Universities, Inc. Basic research in radio astronomy at the U.S. Naval Research Laboratory is supported by 6.1 Base Funding. RJvW acknowledges support from the ERC Starting Grant ClusterWeb 804208. 

\end{acknowledgements}

\vspace{5mm}
\facilities{VLA (NRAO), HST, \textit{Chandra}}

\software{astropy (\cite{astropy}), CASA (\cite{casanew2022}), PyBDSF (\cite{pybdsf}), $\mathrm{TOPCAT}$ (\cite{topcat})}

\clearpage
\bibliography{citation.bib} 

\begin{thebibliography}{}
\expandafter\ifx\csname natexlab\endcsname\relax\def\natexlab#1{#1}\fi
\providecommand{\url}[1]{\href{#1}{#1}}
\providecommand{\dodoi}[1]{doi:~\href{http://doi.org/#1}{\nolinkurl{#1}}}
\providecommand{\doeprint}[1]{\href{http://ascl.net/#1}{\nolinkurl{http://ascl.net/#1}}}
\providecommand{\doarXiv}[1]{\href{https://arxiv.org/abs/#1}{\nolinkurl{https://arxiv.org/abs/#1}}}

\bibitem[{{Akamatsu} {et~al.}(2015){Akamatsu}, {van Weeren}, {Ogrean}, {Kawahara}, {Stroe}, {Sobral}, {Hoeft}, {R{\"o}ttgering}, {Br{\"u}ggen}, \& {Kaastra}}]{akamatsu2015}
{Akamatsu}, H., {van Weeren}, R.~J., {Ogrean}, G.~A., {et~al.} 2015, \aap, 582, A87, \dodoi{10.1051/0004-6361/201425209}

\bibitem[{{Astropy Collaboration} {et~al.}(2018){Astropy Collaboration}, {Price-Whelan}, {Sip{\H{o}}cz}, {G{\"u}nther}, {Lim}, {Crawford}, {Conseil}, {Shupe}, {Craig}, {Dencheva}, {Ginsburg}, {VanderPlas}, {Bradley}, {P{\'e}rez-Su{\'a}rez}, {de Val-Borro}, {Aldcroft}, {Cruz}, {Robitaille}, {Tollerud}, {Ardelean}, {Babej}, {Bach}, {Bachetti}, {Bakanov}, {Bamford}, {Barentsen}, {Barmby}, {Baumbach}, {Berry}, {Biscani}, {Boquien}, {Bostroem}, {Bouma}, {Brammer}, {Bray}, {Breytenbach}, {Buddelmeijer}, {Burke}, {Calderone}, {Cano Rodr{\'\i}guez}, {Cara}, {Cardoso}, {Cheedella}, {Copin}, {Corrales}, {Crichton}, {D'Avella}, {Deil}, {Depagne}, {Dietrich}, {Donath}, {Droettboom}, {Earl}, {Erben}, {Fabbro}, {Ferreira}, {Finethy}, {Fox}, {Garrison}, {Gibbons}, {Goldstein}, {Gommers}, {Greco}, {Greenfield}, {Groener}, {Grollier}, {Hagen}, {Hirst}, {Homeier}, {Horton}, {Hosseinzadeh}, {Hu}, {Hunkeler}, {Ivezi{\'c}}, {Jain}, {Jenness}, {Kanarek}, {Kendrew}, {Kern}, {Kerzendorf}, {Khvalko}, {King}, {Kirkby}, {Kulkarni},
  {Kumar}, {Lee}, {Lenz}, {Littlefair}, {Ma}, {Macleod}, {Mastropietro}, {McCully}, {Montagnac}, {Morris}, {Mueller}, {Mumford}, {Muna}, {Murphy}, {Nelson}, {Nguyen}, {Ninan}, {N{\"o}the}, {Ogaz}, {Oh}, {Parejko}, {Parley}, {Pascual}, {Patil}, {Patil}, {Plunkett}, {Prochaska}, {Rastogi}, {Reddy Janga}, {Sabater}, {Sakurikar}, {Seifert}, {Sherbert}, {Sherwood-Taylor}, {Shih}, {Sick}, {Silbiger}, {Singanamalla}, {Singer}, {Sladen}, {Sooley}, {Sornarajah}, {Streicher}, {Teuben}, {Thomas}, {Tremblay}, {Turner}, {Terr{\'o}n}, {van Kerkwijk}, {de la Vega}, {Watkins}, {Weaver}, {Whitmore}, {Woillez}, {Zabalza}, \& {Astropy Contributors}}]{astropy}
{Astropy Collaboration}, {Price-Whelan}, A.~M., {Sip{\H{o}}cz}, B.~M., {et~al.} 2018, \aj, 156, 123, \dodoi{10.3847/1538-3881/aabc4f}

\bibitem[{{Baars} {et~al.}(1977){Baars}, {Genzel}, {Pauliny-Toth}, \& {Witzel}}]{baars1977}
{Baars}, J.~W.~M., {Genzel}, R., {Pauliny-Toth}, I.~I.~K., \& {Witzel}, A. 1977, \aap, 61, 99

\bibitem[{{Balboni} {et~al.}(2024){Balboni}, {Gastaldello}, {Bonafede}, {Botteon}, {Bartalucci}, {Bourdin}, {Brunetti}, {Cassano}, {De Grandi}, {De Luca}, {Ettori}, {Ghizzardi}, {Gitti}, {Iqbal}, {Johnston-Hollitt}, {Lovisari}, {Mazzotta}, {Molendi}, {Pointecouteau}, {Pratt}, {Riva}, {Rossetti}, {Rottgering}, {Sereno}, {van Weeren}, {Venturi}, \& {Veronesi}}]{balboni2024}
{Balboni}, M., {Gastaldello}, F., {Bonafede}, A., {et~al.} 2024, \aap, 686, A5, \dodoi{10.1051/0004-6361/202347965}

\bibitem[{{Blandford} \& {Eichler}(1987)}]{blandfordeichler1987}
{Blandford}, R., \& {Eichler}, D. 1987, \physrep, 154, 1, \dodoi{10.1016/0370-1573(87)90134-7}

\bibitem[{{Blasi} \& {Colafrancesco}(1999)}]{blasicolafrancesco1999}
{Blasi}, P., \& {Colafrancesco}, S. 1999, Astroparticle Physics, 12, 169, \dodoi{10.1016/S0927-6505(99)00079-1}

\bibitem[{{B{\"o}hringer} \& {Werner}(2010)}]{bohringer2010}
{B{\"o}hringer}, H., \& {Werner}, N. 2010, \aapr, 18, 127, \dodoi{10.1007/s00159-009-0023-3}

\bibitem[{{Bonafede} {et~al.}(2014){Bonafede}, {Intema}, {Br{\"u}ggen}, {Girardi}, {Nonino}, {Kantharia}, {van Weeren}, \& {R{\"o}ttgering}}]{bonafede2014}
{Bonafede}, A., {Intema}, H.~T., {Br{\"u}ggen}, M., {et~al.} 2014, \apj, 785, 1, \dodoi{10.1088/0004-637X/785/1/1}

\bibitem[{{Bonafede} {et~al.}(2022){Bonafede}, {Brunetti}, {Rudnick}, {Vazza}, {Bourdin}, {Giovannini}, {Shimwell}, {Zhang}, {Mazzotta}, {Simionescu}, {Biava}, {Bonnassieux}, {Brienza}, {Br{\"u}ggen}, {Rajpurohit}, {Riseley}, {Stuardi}, {Feretti}, {Tasse}, {Botteon}, {Carretti}, {Cassano}, {Cuciti}, {de Gasperin}, {Gastaldello}, {Rossetti}, {Rottgering}, {Venturi}, \& {van Weeren}}]{bonafede2022}
{Bonafede}, A., {Brunetti}, G., {Rudnick}, L., {et~al.} 2022, \apj, 933, 218, \dodoi{10.3847/1538-4357/ac721d}

\bibitem[{{Botteon} {et~al.}(2020){Botteon}, {Brunetti}, {Ryu}, \& {Roh}}]{botteon2020}
{Botteon}, A., {Brunetti}, G., {Ryu}, D., \& {Roh}, S. 2020, \aap, 634, A64, \dodoi{10.1051/0004-6361/201936216}

\bibitem[{{Botteon} {et~al.}(2016{\natexlab{a}}){Botteon}, {Gastaldello}, {Brunetti}, \& {Dallacasa}}]{botteon2016relic115}
{Botteon}, A., {Gastaldello}, F., {Brunetti}, G., \& {Dallacasa}, D. 2016{\natexlab{a}}, \mnras, 460, L84, \dodoi{10.1093/mnrasl/slw082}

\bibitem[{{Botteon} {et~al.}(2016{\natexlab{b}}){Botteon}, {Gastaldello}, {Brunetti}, \& {Dallacasa}}]{botteon2016shock115}
---. 2016{\natexlab{b}}, Galaxies, 4, 68, \dodoi{10.3390/galaxies4040068}

\bibitem[{{Botteon} {et~al.}(2016{\natexlab{c}}){Botteon}, {Gastaldello}, {Brunetti}, \& {Kale}}]{botteon2016elgordorelic}
{Botteon}, A., {Gastaldello}, F., {Brunetti}, G., \& {Kale}, R. 2016{\natexlab{c}}, \mnras, 463, 1534, \dodoi{10.1093/mnras/stw2089}

\bibitem[{{Botteon} {et~al.}(2018){Botteon}, {Shimwell}, {Bonafede}, {Dallacasa}, {Brunetti}, {Mandal}, {van Weeren}, {Br{\"u}ggen}, {Cassano}, {de Gasperin}, {Hoang}, {Hoeft}, {R{\"o}ttgering}, {Savini}, {White}, {Wilber}, \& {Venturi}}]{botteon2018}
{Botteon}, A., {Shimwell}, T.~W., {Bonafede}, A., {et~al.} 2018, \mnras, 478, 885, \dodoi{10.1093/mnras/sty1102}

\bibitem[{{Botteon} {et~al.}(2024){Botteon}, {van Weeren}, {Eckert}, {Gastaldello}, {Markevitch}, {Giacintucci}, {Brunetti}, {Kale}, \& {Venturi}}]{botteon2024}
{Botteon}, A., {van Weeren}, R.~J., {Eckert}, D., {et~al.} 2024, \aap, 690, A222, \dodoi{10.1051/0004-6361/202451293}

\bibitem[{{Brown} \& {Rudnick}(2011)}]{brown2011}
{Brown}, S., \& {Rudnick}, L. 2011, \mnras, 412, 2, \dodoi{10.1111/j.1365-2966.2010.17738.x}

\bibitem[{{Brunetti} {et~al.}(2001{\natexlab{a}}){Brunetti}, {Cappi}, {Setti}, {Feretti}, \& {Harris}}]{brunetti2001ICscatter}
{Brunetti}, G., {Cappi}, M., {Setti}, G., {Feretti}, L., \& {Harris}, D.~E. 2001{\natexlab{a}}, \aap, 372, 755, \dodoi{10.1051/0004-6361:20010484}

\bibitem[{{Brunetti} \& {Jones}(2014)}]{brunettijones2014}
{Brunetti}, G., \& {Jones}, T.~W. 2014, International Journal of Modern Physics D, 23, 1430007, \dodoi{10.1142/S0218271814300079}

\bibitem[{{Brunetti} {et~al.}(2001{\natexlab{b}}){Brunetti}, {Setti}, {Feretti}, \& {Giovannini}}]{brunetti2001reaccel}
{Brunetti}, G., {Setti}, G., {Feretti}, L., \& {Giovannini}, G. 2001{\natexlab{b}}, \mnras, 320, 365, \dodoi{10.1046/j.1365-8711.2001.03978.x}

\bibitem[{{Campitiello} {et~al.}(2024){Campitiello}, {Bonafede}, {Botteon}, {Lovisari}, {Ettori}, {Brunetti}, {Gastaldello}, {Rossetti}, {Cassano}, {Ignesti}, {van Weeren}, {Br{\"u}ggen}, \& {Hoeft}}]{campitiello2024}
{Campitiello}, M.~G., {Bonafede}, A., {Botteon}, A., {et~al.} 2024, \aap, 683, A9, \dodoi{10.1051/0004-6361/202346591}

\bibitem[{{Caprioli} \& {Spitkovsky}(2014)}]{capriolispitkovsky2014}
{Caprioli}, D., \& {Spitkovsky}, A. 2014, \apj, 783, 91, \dodoi{10.1088/0004-637X/783/2/91}

\bibitem[{{CASA Team} {et~al.}(2022){CASA Team}, {Bean}, {Bhatnagar}, {Castro}, {Donovan Meyer}, {Emonts}, {Garcia}, {Garwood}, {Golap}, {Gonzalez Villalba}, {Harris}, {Hayashi}, {Hoskins}, {Hsieh}, {Jagannathan}, {Kawasaki}, {Keimpema}, {Kettenis}, {Lopez}, {Marvil}, {Masters}, {McNichols}, {Mehringer}, {Miel}, {Moellenbrock}, {Montesino}, {Nakazato}, {Ott}, {Petry}, {Pokorny}, {Raba}, {Rau}, {Schiebel}, {Schweighart}, {Sekhar}, {Shimada}, {Small}, {Steeb}, {Sugimoto}, {Suoranta}, {Tsutsumi}, {van Bemmel}, {Verkouter}, {Wells}, {Xiong}, {Szomoru}, {Griffith}, {Glendenning}, \& {Kern}}]{casanew2022}
{CASA Team}, {Bean}, B., {Bhatnagar}, S., {et~al.} 2022, \pasp, 134, 114501, \dodoi{10.1088/1538-3873/ac9642}

\bibitem[{{Cassano} {et~al.}(2013){Cassano}, {Ettori}, {Brunetti}, {Giacintucci}, {Pratt}, {Venturi}, {Kale}, {Dolag}, \& {Markevitch}}]{cassano2013}
{Cassano}, R., {Ettori}, S., {Brunetti}, G., {et~al.} 2013, \apj, 777, 141, \dodoi{10.1088/0004-637X/777/2/141}

\bibitem[{{Condon} {et~al.}(1998){Condon}, {Cotton}, {Greisen}, {Yin}, {Perley}, {Taylor}, \& {Broderick}}]{condon1998}
{Condon}, J.~J., {Cotton}, W.~D., {Greisen}, E.~W., {et~al.} 1998, \aj, 115, 1693, \dodoi{10.1086/300337}

\bibitem[{{Cornwell}(2008)}]{cornwell2008}
{Cornwell}, T.~J. 2008, IEEE Journal of Selected Topics in Signal Processing, 2, 793, \dodoi{10.1109/JSTSP.2008.2006388}

\bibitem[{{Cotton}(2013)}]{obit}
{Cotton}, B. 2013, {Obit: Radio Astronomy Data Handling}, Astrophysics Source Code Library, record ascl:1307.008.
\newblock \doeprint{1307.008}

\bibitem[{{Cuciti} {et~al.}(2021){Cuciti}, {Cassano}, {Brunetti}, {Dallacasa}, {de Gasperin}, {Ettori}, {Giacintucci}, {Kale}, {Pratt}, {van Weeren}, \& {Venturi}}]{cuciti2021}
{Cuciti}, V., {Cassano}, R., {Brunetti}, G., {et~al.} 2021, \aap, 647, A51, \dodoi{10.1051/0004-6361/202039208}

\bibitem[{{de Gasperin} {et~al.}(2015){de Gasperin}, {Ogrean}, {van Weeren}, {Dawson}, {Br{\"u}ggen}, {Bonafede}, \& {Simionescu}}]{degasperin2015}
{de Gasperin}, F., {Ogrean}, G.~A., {van Weeren}, R.~J., {et~al.} 2015, \mnras, 448, 2197, \dodoi{10.1093/mnras/stv129}

\bibitem[{{de Gasperin} {et~al.}(2014){de Gasperin}, {van Weeren}, {Br{\"u}ggen}, {Vazza}, {Bonafede}, \& {Intema}}]{degasperin2014}
{de Gasperin}, F., {van Weeren}, R.~J., {Br{\"u}ggen}, M., {et~al.} 2014, \mnras, 444, 3130, \dodoi{10.1093/mnras/stu1658}

\bibitem[{{de Gasperin} {et~al.}(2022){de Gasperin}, {Rudnick}, {Finoguenov}, {Wittor}, {Akamatsu}, {Br{\"u}ggen}, {Chibueze}, {Clarke}, {Cotton}, {Cuciti}, {Dom{\'\i}nguez-Fern{\'a}ndez}, {Knowles}, {O'Sullivan}, \& {Sebokolodi}}]{degasperin2022}
{de Gasperin}, F., {Rudnick}, L., {Finoguenov}, A., {et~al.} 2022, \aap, 659, A146, \dodoi{10.1051/0004-6361/202142658}

\bibitem[{{Dennison}(1980)}]{dennison1980}
{Dennison}, B. 1980, \apjl, 239, L93, \dodoi{10.1086/183300}

\bibitem[{{Di Gennaro} {et~al.}(2018){Di Gennaro}, {Venturi}, {Dallacasa}, {Giacintucci}, {Merluzzi}, {Busarello}, {Mercurio}, {Bardelli}, {Gastaldello}, {Grado}, {Haines}, {Limatola}, \& {Rossetti}}]{digennaro2018}
{Di Gennaro}, G., {Venturi}, T., {Dallacasa}, D., {et~al.} 2018, \aap, 620, A25, \dodoi{10.1051/0004-6361/201832801}

\bibitem[{{Di Gennaro} {et~al.}(2021){Di Gennaro}, {van Weeren}, {Rudnick}, {Hoeft}, {Br{\"u}ggen}, {Ryu}, {R{\"o}ttgering}, {Forman}, {Stroe}, {Shimwell}, {Kraft}, {Jones}, \& {Hoang}}]{digennaro2021}
{Di Gennaro}, G., {van Weeren}, R.~J., {Rudnick}, L., {et~al.} 2021, \apj, 911, 3, \dodoi{10.3847/1538-4357/abe620}

\bibitem[{{Dominguez-Fernandez} {et~al.}(2021){Dominguez-Fernandez}, {Bruggen}, {Vazza}, {Banda-Barragan}, {Rajpurohit}, {Mignone}, {Mukherjee}, \& {Vaidya}}]{dominguezfernandez2021}
{Dominguez-Fernandez}, P., {Bruggen}, M., {Vazza}, F., {et~al.} 2021, \mnras, 500, 795, \dodoi{10.1093/mnras/staa3018}

\bibitem[{{Dutta} {et~al.}(2023){Dutta}, {Singh}, {Chandra}, {Wadadekar}, {Kayal}, \& {Heywood}}]{dutta2023}
{Dutta}, S., {Singh}, V., {Chandra}, C.~H.~I., {et~al.} 2023, \apj, 944, 176, \dodoi{10.3847/1538-4357/acaf01}

\bibitem[{{Ebeling} {et~al.}(2002){Ebeling}, {Mullis}, \& {Tully}}]{ebeling2002}
{Ebeling}, H., {Mullis}, C.~R., \& {Tully}, R.~B. 2002, \apj, 580, 774, \dodoi{10.1086/343790}

\bibitem[{{Eckert} {et~al.}(2020){Eckert}, {Finoguenov}, {Ghirardini}, {Grandis}, {Kaefer}, {Sanders}, \& {Ramos-Ceja}}]{eckert2020}
{Eckert}, D., {Finoguenov}, A., {Ghirardini}, V., {et~al.} 2020, The Open Journal of Astrophysics, 3, 12, \dodoi{10.21105/astro.2009.13944}

\bibitem[{{Ensslin} {et~al.}(1998){Ensslin}, {Biermann}, {Klein}, \& {Kohle}}]{ensslin1998relics}
{Ensslin}, T.~A., {Biermann}, P.~L., {Klein}, U., \& {Kohle}, S. 1998, \aap, 332, 395, \dodoi{10.48550/arXiv.astro-ph/9712293}

\bibitem[{{En{\ss}lin} \& {Gopal-Krishna}(2001)}]{ensslin2001}
{En{\ss}lin}, T.~A., \& {Gopal-Krishna}. 2001, \aap, 366, 26, \dodoi{10.1051/0004-6361:20000198}

\bibitem[{{Feretti} {et~al.}(2012){Feretti}, {Giovannini}, {Govoni}, \& {Murgia}}]{feretti2012}
{Feretti}, L., {Giovannini}, G., {Govoni}, F., \& {Murgia}, M. 2012, \aapr, 20, 54, \dodoi{10.1007/s00159-012-0054-z}

\bibitem[{{Fermi}(1949)}]{fermi1949}
{Fermi}, E. 1949, Physical Review, 75, 1169, \dodoi{10.1103/PhysRev.75.1169}

\bibitem[{{Finner} {et~al.}(2023){Finner}, {Randall}, {Jee}, {Blanton}, {Cho}, {Clarke}, {Giacintucci}, {Nulsen}, \& {van Weeren}}]{Finner_2023}
{Finner}, K., {Randall}, S.~W., {Jee}, M.~J., {et~al.} 2023, \apj, 942, 23, \dodoi{10.3847/1538-4357/ac9fd3}

\bibitem[{{Forman} \& {Jones}(1982)}]{forman1982}
{Forman}, W., \& {Jones}, C. 1982, \araa, 20, 547, \dodoi{10.1146/annurev.aa.20.090182.002555}

\bibitem[{{Giacintucci} {et~al.}(2008){Giacintucci}, {Venturi}, {Macario}, {Dallacasa}, {Brunetti}, {Markevitch}, {Cassano}, {Bardelli}, \& {Athreya}}]{giacintucci2008}
{Giacintucci}, S., {Venturi}, T., {Macario}, G., {et~al.} 2008, \aap, 486, 347, \dodoi{10.1051/0004-6361:200809459}

\bibitem[{{Govoni} {et~al.}(2001{\natexlab{a}}){Govoni}, {En{\ss}lin}, {Feretti}, \& {Giovannini}}]{govoni2001}
{Govoni}, F., {En{\ss}lin}, T.~A., {Feretti}, L., \& {Giovannini}, G. 2001{\natexlab{a}}, \aap, 369, 441, \dodoi{10.1051/0004-6361:20010115}

\bibitem[{{Govoni} {et~al.}(2001{\natexlab{b}}){Govoni}, {Feretti}, {Giovannini}, {B{\"o}hringer}, {Reiprich}, \& {Murgia}}]{govoni2001b}
{Govoni}, F., {Feretti}, L., {Giovannini}, G., {et~al.} 2001{\natexlab{b}}, \aap, 376, 803, \dodoi{10.1051/0004-6361:20011016}

\bibitem[{{Greisen}(2003)}]{aipsgreisen}
{Greisen}, E.~W. 2003, in Astrophysics and Space Science Library, Vol. 285, Information Handling in Astronomy - Historical Vistas, ed. A.~{Heck}, 109, \dodoi{10.1007/0-306-48080-8_7}

\bibitem[{{Hoang} {et~al.}(2018){Hoang}, {Shimwell}, {van Weeren}, {Intema}, {R{\"o}ttgering}, {Andrade-Santos}, {Akamatsu}, {Bonafede}, {Brunetti}, {Dawson}, {Golovich}, {Best}, {Botteon}, {Br{\"u}ggen}, {Cassano}, {de Gasperin}, {Hoeft}, {Stroe}, \& {White}}]{hoang2018}
{Hoang}, D.~N., {Shimwell}, T.~W., {van Weeren}, R.~J., {et~al.} 2018, \mnras, 478, 2218, \dodoi{10.1093/mnras/sty1123}

\bibitem[{{Hoeft} \& {Br{\"u}ggen}(2007)}]{hoeftbruggen2007}
{Hoeft}, M., \& {Br{\"u}ggen}, M. 2007, \mnras, 375, 77, \dodoi{10.1111/j.1365-2966.2006.11111.x}

\bibitem[{{Jones} {et~al.}(2023){Jones}, {de Gasperin}, {Cuciti}, {Botteon}, {Zhang}, {Gastaldello}, {Shimwell}, {Simionescu}, {Rossetti}, {Cassano}, {Akamatsu}, {Bonafede}, {Br{\"u}ggen}, {Brunetti}, {Camillini}, {Di Gennaro}, {Drabent}, {Hoang}, {Rajpurohit}, {Natale}, {Tasse}, \& {van Weeren}}]{jones2023}
{Jones}, A., {de Gasperin}, F., {Cuciti}, V., {et~al.} 2023, \aap, 680, A31, \dodoi{10.1051/0004-6361/202245102}

\bibitem[{{Kale} \& {Dwarakanath}(2012)}]{kale2012}
{Kale}, R., \& {Dwarakanath}, K.~S. 2012, \apj, 744, 46, \dodoi{10.1088/0004-637X/744/1/46}

\bibitem[{{Kang} {et~al.}(2012){Kang}, {Ryu}, \& {Jones}}]{kang2012}
{Kang}, H., {Ryu}, D., \& {Jones}, T.~W. 2012, \apj, 756, 97, \dodoi{10.1088/0004-637X/756/1/97}

\bibitem[{{Katz-Stone} \& {Rudnick}(1997)}]{katzstone1997}
{Katz-Stone}, D.~M., \& {Rudnick}, L. 1997, \apj, 488, 146, \dodoi{10.1086/304661}

\bibitem[{{Katz-Stone} {et~al.}(1993){Katz-Stone}, {Rudnick}, \& {Anderson}}]{katzstone1993}
{Katz-Stone}, D.~M., {Rudnick}, L., \& {Anderson}, M.~C. 1993, \apj, 407, 549, \dodoi{10.1086/172536}

\bibitem[{{Kelly}(2007)}]{kelly2007}
{Kelly}, B.~C. 2007, \apj, 665, 1489, \dodoi{10.1086/519947}

\bibitem[{{Kempner} {et~al.}(2004){Kempner}, {Blanton}, {Clarke}, {En{\ss}lin}, {Johnston-Hollitt}, \& {Rudnick}}]{kempner2004}
{Kempner}, J.~C., {Blanton}, E.~L., {Clarke}, T.~E., {et~al.} 2004, in The Riddle of Cooling Flows in Galaxies and Clusters of galaxies, ed. T.~{Reiprich}, J.~{Kempner}, \& N.~{Soker}, 335, \dodoi{10.48550/arXiv.astro-ph/0310263}

\bibitem[{Kocevski {et~al.}(2003)Kocevski, Ebeling, \& Mullis}]{kocevski2003clusters}
Kocevski, D.~D., Ebeling, H., \& Mullis, C.~R. 2003, Clusters in the Zone of Avoidance.
\newblock \doarXiv{astro-ph/0304453}

\bibitem[{Kravtsov \& Borgani(2012)}]{Kravtsov_2012}
Kravtsov, A.~V., \& Borgani, S. 2012, Annual Review of Astronomy and Astrophysics, 50, 353, \dodoi{10.1146/annurev-astro-081811-125502}

\bibitem[{{Landau} \& {Lifshitz}(1959)}]{landau1959}
{Landau}, L.~D., \& {Lifshitz}, E.~M. 1959, {Fluid mechanics}

\bibitem[{{Lane} {et~al.}(2014){Lane}, {Cotton}, {van Velzen}, {Clarke}, {Kassim}, {Helmboldt}, {Lazio}, \& {Cohen}}]{lane2014}
{Lane}, W.~M., {Cotton}, W.~D., {van Velzen}, S., {et~al.} 2014, \mnras, 440, 327, \dodoi{10.1093/mnras/stu256}

\bibitem[{{Lee} {et~al.}(2024){Lee}, {Pillepich}, {ZuHone}, {Nelson}, {Jee}, {Nagai}, \& {Finner}}]{wonki2024}
{Lee}, W., {Pillepich}, A., {ZuHone}, J., {et~al.} 2024, \aap, 686, A55, \dodoi{10.1051/0004-6361/202348194}

\bibitem[{{Mandal} {et~al.}(2020){Mandal}, {Intema}, {van Weeren}, {Shimwell}, {Botteon}, {Brunetti}, {de Gasperin}, {Br{\"u}ggen}, {Di Gennaro}, {Kraft}, {R{\"o}ttgering}, {Hardcastle}, \& {Tasse}}]{mandal2020}
{Mandal}, S., {Intema}, H.~T., {van Weeren}, R.~J., {et~al.} 2020, \aap, 634, A4, \dodoi{10.1051/0004-6361/201936560}

\bibitem[{{Markevitch} {et~al.}(2005{\natexlab{a}}){Markevitch}, {Govoni}, {Brunetti}, \& {Jerius}}]{markevitch2005}
{Markevitch}, M., {Govoni}, F., {Brunetti}, G., \& {Jerius}, D. 2005{\natexlab{a}}, \apj, 627, 733, \dodoi{10.1086/430695}

\bibitem[{{Markevitch} {et~al.}(2005{\natexlab{b}}){Markevitch}, {Govoni}, {Brunetti}, \& {Jerius}}]{markevitch520}
---. 2005{\natexlab{b}}, \apj, 627, 733, \dodoi{10.1086/430695}

\bibitem[{Markevitch \& Vikhlinin(2007)}]{Markevitch_2007}
Markevitch, M., \& Vikhlinin, A. 2007, Physics Reports, 443, 1, \dodoi{10.1016/j.physrep.2007.01.001}

\bibitem[{{McDonald} {et~al.}(2022){McDonald}, {Obreschkow}, \& {Garratt-Smithson}}]{mcdonald2022}
{McDonald}, W., {Obreschkow}, D., \& {Garratt-Smithson}, L. 2022, \mnras, 516, 5289, \dodoi{10.1093/mnras/stac2276}

\bibitem[{{Mohan} \& {Rafferty}(2015)}]{pybdsf}
{Mohan}, N., \& {Rafferty}, D. 2015, {PyBDSF: Python Blob Detection and Source Finder}, Astrophysics Source Code Library, record ascl:1502.007.
\newblock \doeprint{1502.007}

\bibitem[{{Murgia} {et~al.}(2010){Murgia}, {Govoni}, {Feretti}, \& {Giovannini}}]{murgia2010}
{Murgia}, M., {Govoni}, F., {Feretti}, L., \& {Giovannini}, G. 2010, \aap, 509, A86, \dodoi{10.1051/0004-6361/200913414}

\bibitem[{{Murgia} {et~al.}(2005){Murgia}, {Parma}, {de Ruiter}, {Mack}, \& {Fanti}}]{murgia2005}
{Murgia}, M., {Parma}, P., {de Ruiter}, H.~R., {Mack}, K.~H., \& {Fanti}, R. 2005, in X-Ray and Radio Connections, ed. L.~O. {Sjouwerman} \& K.~K. {Dyer}, 8.15, \dodoi{10.48550/arXiv.astro-ph/0405091}

\bibitem[{{National Radio Astronomy Observatory (NRAO)}(2024{\natexlab{a}})}]{CASA_PBandGuide}
{National Radio Astronomy Observatory (NRAO)}. 2024{\natexlab{a}}, P-band CASA Guide 5.0.
\newblock \url{https://casaguides.nrao.edu/index.php?title=CASA_Guides:P-band_casa_guide_5.0}

\bibitem[{{National Radio Astronomy Observatory (NRAO)}(2024{\natexlab{b}})}]{NRAO_FluxDensityScale}
---. 2024{\natexlab{b}}, Flux Density Scale.
\newblock \url{https://science.nrao.edu/facilities/vla/docs/manuals/oss/performance/fdscale}

\bibitem[{{Offringa} {et~al.}(2014){Offringa}, {McKinley}, {Hurley-Walker}, {Briggs}, {Wayth}, {Kaplan}, {Bell}, {Feng}, {Neben}, {Hughes}, {Rhee}, {Murphy}, {Bhat}, {Bernardi}, {Bowman}, {Cappallo}, {Corey}, {Deshpande}, {Emrich}, {Ewall-Wice}, {Gaensler}, {Goeke}, {Greenhill}, {Hazelton}, {Hindson}, {Johnston-Hollitt}, {Jacobs}, {Kasper}, {Kratzenberg}, {Lenc}, {Lonsdale}, {Lynch}, {McWhirter}, {Mitchell}, {Morales}, {Morgan}, {Kudryavtseva}, {Oberoi}, {Ord}, {Pindor}, {Procopio}, {Prabu}, {Riding}, {Roshi}, {Shankar}, {Srivani}, {Subrahmanyan}, {Tingay}, {Waterson}, {Webster}, {Whitney}, {Williams}, \& {Williams}}]{wsclean}
{Offringa}, A.~R., {McKinley}, B., {Hurley-Walker}, N., {et~al.} 2014, \mnras, 444, 606, \dodoi{10.1093/mnras/stu1368}

\bibitem[{{Pal} {et~al.}(2024){Pal}, {Kale}, {Wang}, \& {Wik}}]{pal2024}
{Pal}, A., {Kale}, R., {Wang}, Q. H.~S., \& {Wik}, D.~R. 2024, arXiv e-prints, arXiv:2411.15480, \dodoi{10.48550/arXiv.2411.15480}

\bibitem[{{Patil} {et~al.}(2021){Patil}, {Whittle}, {Nyland}, {Lonsdale}, {Lacy}, {Kimball}, {Lonsdale}, {Peters}, {Clarke}, {Efstathiou}, {Giacintucci}, {Kim}, {Lanz}, {Mukherjee}, \& {Polisensky}}]{patil2021}
{Patil}, P., {Whittle}, M., {Nyland}, K., {et~al.} 2021, Astronomische Nachrichten, 342, 1166, \dodoi{10.1002/asna.20210078}

\bibitem[{{Patil} {et~al.}(2022){Patil}, {Whittle}, {Nyland}, {Lonsdale}, {Lacy}, {Kimball}, {Lonsdale}, {Peters}, {Clarke}, {Efstathiou}, {Giacintucci}, {Kim}, {Lanz}, {Mukherjee}, \& {Polisensky}}]{patil2022}
---. 2022, \apj, 934, 26, \dodoi{10.3847/1538-4357/ac71b0}

\bibitem[{{Pearce} {et~al.}(2017){Pearce}, {van Weeren}, {Andrade-Santos}, {Jones}, {Forman}, {Br{\"u}ggen}, {Bulbul}, {Clarke}, {Kraft}, {Medezinski}, {Mroczkowski}, {Nonino}, {Nulsen}, {Randall}, \& {Umetsu}}]{pearce2017}
{Pearce}, C.~J.~J., {van Weeren}, R.~J., {Andrade-Santos}, F., {et~al.} 2017, \apj, 845, 81, \dodoi{10.3847/1538-4357/aa7e2f}

\bibitem[{{Perley} \& {Butler}(2017)}]{perley2017}
{Perley}, R.~A., \& {Butler}, B.~J. 2017, \apjs, 230, 7, \dodoi{10.3847/1538-4365/aa6df9}

\bibitem[{{Petrosian}(2001)}]{petrosian2001}
{Petrosian}, V. 2001, \apj, 557, 560, \dodoi{10.1086/321557}

\bibitem[{{Pfrommer} \& {En{\ss}lin}(2004)}]{pfrommerensslin2004}
{Pfrommer}, C., \& {En{\ss}lin}, T.~A. 2004, Journal of Korean Astronomical Society, 37, 455, \dodoi{10.5303/JKAS.2004.37.5.455}

\bibitem[{{Pinzke} {et~al.}(2013){Pinzke}, {Oh}, \& {Pfrommer}}]{pinzke2013}
{Pinzke}, A., {Oh}, S.~P., \& {Pfrommer}, C. 2013, \mnras, 435, 1061, \dodoi{10.1093/mnras/stt1308}

\bibitem[{{Planck Collaboration} {et~al.}(2020){Planck Collaboration}, {Aghanim}, {Akrami}, {Ashdown}, {Aumont}, {Baccigalupi}, {Ballardini}, {Banday}, {Barreiro}, {Bartolo}, {Basak}, {Battye}, {Benabed}, {Bernard}, {Bersanelli}, {Bielewicz}, {Bock}, {Bond}, {Borrill}, {Bouchet}, {Boulanger}, {Bucher}, {Burigana}, {Butler}, {Calabrese}, {Cardoso}, {Carron}, {Challinor}, {Chiang}, {Chluba}, {Colombo}, {Combet}, {Contreras}, {Crill}, {Cuttaia}, {de Bernardis}, {de Zotti}, {Delabrouille}, {Delouis}, {Di Valentino}, {Diego}, {Dor{\'e}}, {Douspis}, {Ducout}, {Dupac}, {Dusini}, {Efstathiou}, {Elsner}, {En{\ss}lin}, {Eriksen}, {Fantaye}, {Farhang}, {Fergusson}, {Fernandez-Cobos}, {Finelli}, {Forastieri}, {Frailis}, {Fraisse}, {Franceschi}, {Frolov}, {Galeotta}, {Galli}, {Ganga}, {G{\'e}nova-Santos}, {Gerbino}, {Ghosh}, {Gonz{\'a}lez-Nuevo}, {G{\'o}rski}, {Gratton}, {Gruppuso}, {Gudmundsson}, {Hamann}, {Handley}, {Hansen}, {Herranz}, {Hildebrandt}, {Hivon}, {Huang}, {Jaffe}, {Jones}, {Karakci}, {Keih{\"a}nen},
  {Keskitalo}, {Kiiveri}, {Kim}, {Kisner}, {Knox}, {Krachmalnicoff}, {Kunz}, {Kurki-Suonio}, {Lagache}, {Lamarre}, {Lasenby}, {Lattanzi}, {Lawrence}, {Le Jeune}, {Lemos}, {Lesgourgues}, {Levrier}, {Lewis}, {Liguori}, {Lilje}, {Lilley}, {Lindholm}, {L{\'o}pez-Caniego}, {Lubin}, {Ma}, {Mac{\'\i}as-P{\'e}rez}, {Maggio}, {Maino}, {Mandolesi}, {Mangilli}, {Marcos-Caballero}, {Maris}, {Martin}, {Martinelli}, {Mart{\'\i}nez-Gonz{\'a}lez}, {Matarrese}, {Mauri}, {McEwen}, {Meinhold}, {Melchiorri}, {Mennella}, {Migliaccio}, {Millea}, {Mitra}, {Miville-Desch{\^e}nes}, {Molinari}, {Montier}, {Morgante}, {Moss}, {Natoli}, {N{\o}rgaard-Nielsen}, {Pagano}, {Paoletti}, {Partridge}, {Patanchon}, {Peiris}, {Perrotta}, {Pettorino}, {Piacentini}, {Polastri}, {Polenta}, {Puget}, {Rachen}, {Reinecke}, {Remazeilles}, {Renzi}, {Rocha}, {Rosset}, {Roudier}, {Rubi{\~n}o-Mart{\'\i}n}, {Ruiz-Granados}, {Salvati}, {Sandri}, {Savelainen}, {Scott}, {Shellard}, {Sirignano}, {Sirri}, {Spencer}, {Sunyaev}, {Suur-Uski}, {Tauber}, {Tavagnacco},
  {Tenti}, {Toffolatti}, {Tomasi}, {Trombetti}, {Valenziano}, {Valiviita}, {Van Tent}, {Vibert}, {Vielva}, {Villa}, {Vittorio}, {Wandelt}, {Wehus}, {White}, {White}, {Zacchei}, \& {Zonca}}]{planck2020}
{Planck Collaboration}, {Aghanim}, N., {Akrami}, Y., {et~al.} 2020, \aap, 641, A6, \dodoi{10.1051/0004-6361/201833910}

\bibitem[{Planelles \& Quilis(2009)}]{Planelles_2009}
Planelles, S., \& Quilis, V. 2009, Monthly Notices of the Royal Astronomical Society, 399, 410, \dodoi{10.1111/j.1365-2966.2009.15290.x}

\bibitem[{{Rajpurohit} {et~al.}(2018){Rajpurohit}, {Hoeft}, {van Weeren}, {Rudnick}, {R{\"o}ttgering}, {Forman}, {Br{\"u}ggen}, {Croston}, {Andrade-Santos}, {Dawson}, {Intema}, {Kraft}, {Jones}, \& {Jee}}]{rajpurohit2018}
{Rajpurohit}, K., {Hoeft}, M., {van Weeren}, R.~J., {et~al.} 2018, \apj, 852, 65, \dodoi{10.3847/1538-4357/aa9f13}

\bibitem[{{Rajpurohit} {et~al.}(2020){Rajpurohit}, {Vazza}, {Hoeft}, {Loi}, {Beck}, {Vacca}, {Kierdorf}, {van Weeren}, {Wittor}, {Govoni}, {Murgia}, {Riseley}, {Locatelli}, {Drabent}, \& {Bonnassieux}}]{rajpurohit2020}
{Rajpurohit}, K., {Vazza}, F., {Hoeft}, M., {et~al.} 2020, \aap, 642, L13, \dodoi{10.1051/0004-6361/202039165}

\bibitem[{{Rajpurohit} {et~al.}(2023){Rajpurohit}, {Osinga}, {Brienza}, {Botteon}, {Brunetti}, {Forman}, {Riseley}, {Vazza}, {Bonafede}, {van Weeren}, {Br{\"u}ggen}, {Rajpurohit}, {Drabent}, {Dallacasa}, {Rossetti}, {Rajpurohit}, {Hoeft}, {Bonnassieux}, {Cassano}, \& {Miley}}]{rajpurohit2023}
{Rajpurohit}, K., {Osinga}, E., {Brienza}, M., {et~al.} 2023, \aap, 669, A1, \dodoi{10.1051/0004-6361/202244925}

\bibitem[{Randall {et~al.}(2016)Randall, Clarke, van Weeren, Intema, Dawson, Mroczkowski, Blanton, Bulbul, \& Giacintucci}]{Randall_2016}
Randall, S.~W., Clarke, T.~E., van Weeren, R.~J., {et~al.} 2016, The Astrophysical Journal, 823, 94, \dodoi{10.3847/0004-637x/823/2/94}

\bibitem[{{Sarazin}(1988)}]{sarazin1988}
{Sarazin}, C.~L. 1988, {X-ray emission from clusters of galaxies}

\bibitem[{{Sarazin}(2002)}]{sarazin2002}
{Sarazin}, C.~L. 2002, in Astrophysics and Space Science Library, Vol. 272, Merging Processes in Galaxy Clusters, ed. L.~{Feretti}, I.~M. {Gioia}, \& G.~{Giovannini}, 1--38, \dodoi{10.1007/0-306-48096-4_1}

\bibitem[{{Sarazin} \& {Kempner}(2000)}]{sarazin2000}
{Sarazin}, C.~L., \& {Kempner}, J.~C. 2000, \apj, 533, 73, \dodoi{10.1086/308649}

\bibitem[{{Sarkar} {et~al.}(2022){Sarkar}, {Randall}, {Su}, {Alvarez}, {Sarazin}, {Nulsen}, {Blanton}, {Forman}, {Jones}, {Bulbul}, {Zuhone}, {Andrade-Santos}, {Johnson}, \& {Chakraborty}}]{2022ApJ...935L..23S}
{Sarkar}, A., {Randall}, S., {Su}, Y., {et~al.} 2022, \apjl, 935, L23, \dodoi{10.3847/2041-8213/ac86d4}

\bibitem[{{Shimwell} {et~al.}(2014){Shimwell}, {Brown}, {Feain}, {Feretti}, {Gaensler}, \& {Lage}}]{shimwell2014}
{Shimwell}, T.~W., {Brown}, S., {Feain}, I.~J., {et~al.} 2014, \mnras, 440, 2901, \dodoi{10.1093/mnras/stu467}

\bibitem[{{Taylor}(2005)}]{topcat}
{Taylor}, M.~B. 2005, in Astronomical Society of the Pacific Conference Series, Vol. 347, Astronomical Data Analysis Software and Systems XIV, ed. P.~{Shopbell}, M.~{Britton}, \& R.~{Ebert}, 29

\bibitem[{{Tormen} {et~al.}(2004){Tormen}, {Moscardini}, \& {Yoshida}}]{tormen2004}
{Tormen}, G., {Moscardini}, L., \& {Yoshida}, N. 2004, \mnras, 350, 1397, \dodoi{10.1111/j.1365-2966.2004.07736.x}

\bibitem[{{Umetsu} {et~al.}(2020){Umetsu}, {Sereno}, {Lieu}, {Miyatake}, {Medezinski}, {Nishizawa}, {Giles}, {Gastaldello}, {McCarthy}, {Kilbinger}, {Birkinshaw}, {Ettori}, {Okabe}, {Chiu}, {Coupon}, {Eckert}, {Fujita}, {Higuchi}, {Koulouridis}, {Maughan}, {Miyazaki}, {Oguri}, {Pacaud}, {Pierre}, {Rapetti}, \& {Smith}}]{umetsu2020}
{Umetsu}, K., {Sereno}, M., {Lieu}, M., {et~al.} 2020, \apj, 890, 148, \dodoi{10.3847/1538-4357/ab6bca}

\bibitem[{{van Weeren} {et~al.}(2011{\natexlab{a}}){van Weeren}, {Br{\"u}ggen}, {R{\"o}ttgering}, \& {Hoeft}}]{vanweeren2011relics}
{van Weeren}, R.~J., {Br{\"u}ggen}, M., {R{\"o}ttgering}, H.~J.~A., \& {Hoeft}, M. 2011{\natexlab{a}}, Journal of Astrophysics and Astronomy, 32, 505, \dodoi{10.1007/s12036-011-9108-2}

\bibitem[{{van Weeren} {et~al.}(2011{\natexlab{b}}){van Weeren}, {Br{\"u}ggen}, {R{\"o}ttgering}, {Hoeft}, {Nuza}, \& {Intema}}]{vanweeren2011}
{van Weeren}, R.~J., {Br{\"u}ggen}, M., {R{\"o}ttgering}, H.~J.~A., {et~al.} 2011{\natexlab{b}}, \aap, 533, A35, \dodoi{10.1051/0004-6361/201117149}

\bibitem[{{van Weeren} {et~al.}(2019){van Weeren}, {de Gasperin}, {Akamatsu}, {Br{\"u}ggen}, {Feretti}, {Kang}, {Stroe}, \& {Zandanel}}]{vanweerenreview}
{van Weeren}, R.~J., {de Gasperin}, F., {Akamatsu}, H., {et~al.} 2019, \ssr, 215, 16, \dodoi{10.1007/s11214-019-0584-z}

\bibitem[{van Weeren {et~al.}(2019)van Weeren, de~Gasperin, Akamatsu, Brüggen, Feretti, Kang, Stroe, \& Zandanel}]{van_Weeren_2019}
van Weeren, R.~J., de~Gasperin, F., Akamatsu, H., {et~al.} 2019, Space Science Reviews, 215, \dodoi{10.1007/s11214-019-0584-z}

\bibitem[{{van Weeren} {et~al.}(2011{\natexlab{c}}){van Weeren}, {R{\"o}ttgering}, \& {Br{\"u}ggen}}]{vanweeren2011a}
{van Weeren}, R.~J., {R{\"o}ttgering}, H.~J.~A., \& {Br{\"u}ggen}, M. 2011{\natexlab{c}}, \aap, 527, A114, \dodoi{10.1051/0004-6361/201015991}

\bibitem[{{van Weeren} {et~al.}(2016){van Weeren}, {Brunetti}, {Br{\"u}ggen}, {Andrade-Santos}, {Ogrean}, {Williams}, {R{\"o}ttgering}, {Dawson}, {Forman}, {de Gasperin}, {Hardcastle}, {Jones}, {Miley}, {Rafferty}, {Rudnick}, {Sabater}, {Sarazin}, {Shimwell}, {Bonafede}, {Best}, {B{\^\i}rzan}, {Cassano}, {Chy{\.z}y}, {Croston}, {Dijkema}, {En{\ss}lin}, {Ferrari}, {Heald}, {Hoeft}, {Horellou}, {Jarvis}, {Kraft}, {Mevius}, {Intema}, {Murray}, {Orr{\'u}}, {Pizzo}, {Sridhar}, {Simionescu}, {Stroe}, {van der Tol}, \& {White}}]{vanweeren2016}
{van Weeren}, R.~J., {Brunetti}, G., {Br{\"u}ggen}, M., {et~al.} 2016, \apj, 818, 204, \dodoi{10.3847/0004-637X/818/2/204}

\bibitem[{{van Weeren} {et~al.}(2017){van Weeren}, {Andrade-Santos}, {Dawson}, {Golovich}, {Lal}, {Kang}, {Ryu}, {Br{\`\i}ggen}, {Ogrean}, {Forman}, {Jones}, {Placco}, {Santucci}, {Wittman}, {Jee}, {Kraft}, {Sobral}, {Stroe}, \& {Fogarty}}]{vanweeren2017accel}
{van Weeren}, R.~J., {Andrade-Santos}, F., {Dawson}, W.~A., {et~al.} 2017, Nature Astronomy, 1, 0005, \dodoi{10.1038/s41550-016-0005}

\bibitem[{{Wang} {et~al.}(2018){Wang}, {Giacintucci}, \& {Markevitch}}]{wang2018}
{Wang}, Q. H.~S., {Giacintucci}, S., \& {Markevitch}, M. 2018, \apj, 856, 162, \dodoi{10.3847/1538-4357/aab2aa}

\bibitem[{ZuHone \& Su(2012)}]{ZuHone_2012}
ZuHone, J., \& Su, Y. 2012, The Merger Dynamics of the X-Ray-Emitting Plasma in Clusters of Galaxies (Springer Nature Singapore), 1–44, \dodoi{10.1007/978-981-16-4544-0_124-1}

\end{thebibliography}

\bibliographystyle{aasjournal}

\section{Appendix A: Point Source Subtraction}
\label{sec:appendixA}

CIZA J0107.7+5408 (CIZA0107) is part of the X-ray survey for Clusters in the Zone of Avoidance \citep[CIZA;][]{kocevski2003clusters}. The CIZA catalog is the first statistically complete catalog of X-ray selected galaxy clusters behind the Galactic plane. The field is highly contaminated with compact emission, and thus in order to fully constrain the diffuse emission of the cluster merger, it was necessary to subtract off the contaminant point sources.

\begin{figure*}[ht!]
\centering
\begin{minipage}{0.45\textwidth}
    \includegraphics[width=\linewidth]{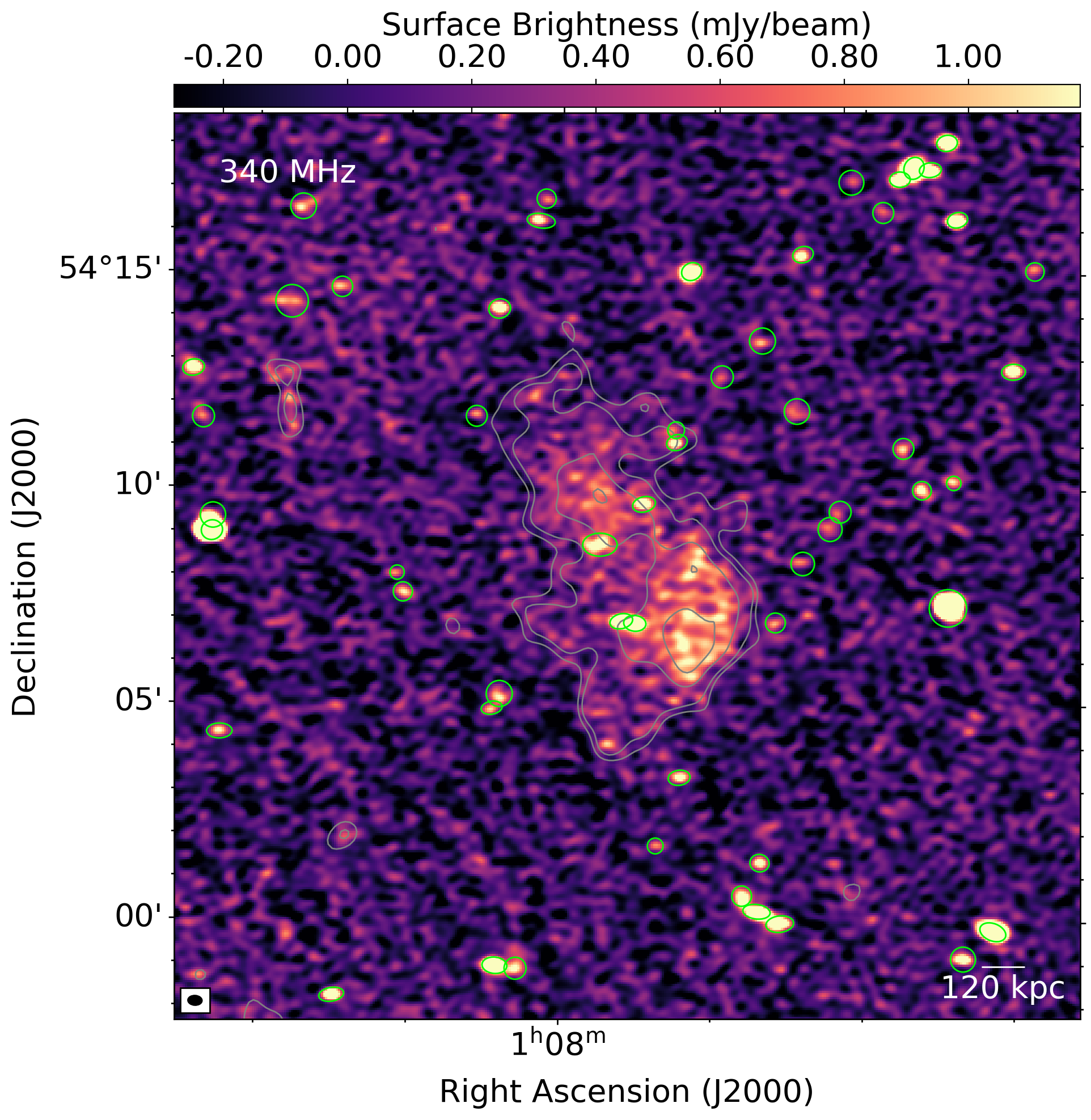}
\end{minipage}
\begin{minipage}{0.45\textwidth}
    \includegraphics[width=\linewidth]{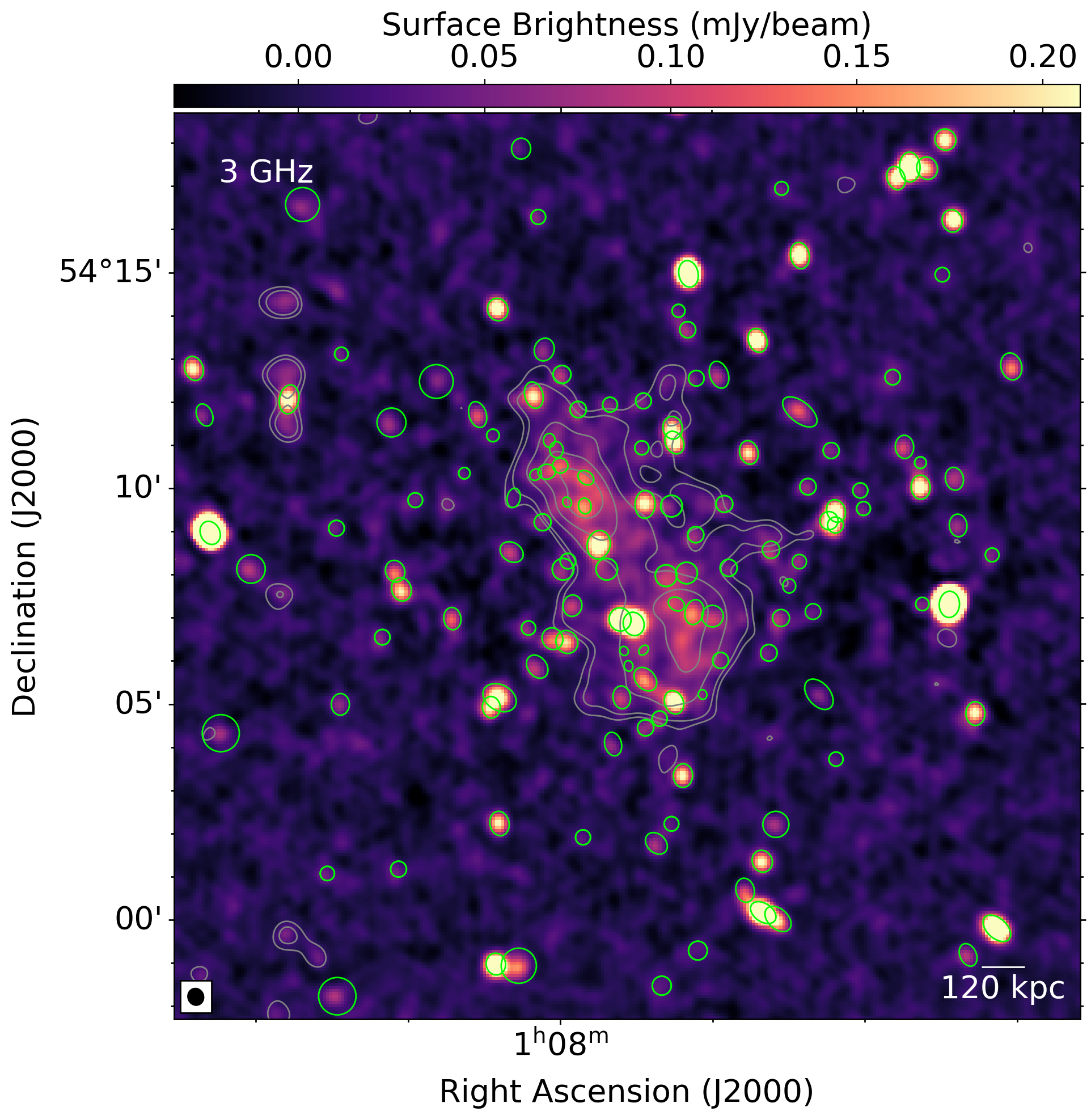}
\end{minipage}
\caption{A (left): 340 MHz VLA observations at B-configuration with a beam of 19.44'' $\times{}$ 13.28'' $\times{}$ 86$^{\circ}$, marked with all regions used in point source subtraction. To give a sense of the location of the radio structure, contours are displayed tracing the 340 MHz emission convolved to a 40'' beam, beginning at 3$\sigma$ and proceeding in integer multiples of $\sqrt{2}$. B (right): 3 GHz VLA observations at D-configuration with a beam 23.33'' $\times{}$ 21.46'' $\times{}$ 20$^{\circ}$, marked with all regions used in point source subtraction. Contours are displayed tracing the 3 GHz emission convolved to a 40'' beam, beginning at 3$\sigma$ and proceeding in integer multiples of $\sqrt{2}$. The physical scale is 1.95 kiloparsecs per arcsecond.}
\label{fig:radio_pointsources}
\end{figure*}

The compact emission was subtracted from both the P-band, B configuration observations and the S-band, D configuration observations. Figures \ref{fig:ciza_nativeims}B and \ref{fig:ciza_nativeims}D show the pre-subtraction continuum images, and it is clear how significantly the halo is impacted by the compact emission. For each band, the fully calibrated visibilities were cleaned with \texttt{WSClean}, with an inner cut on the uv-data to ensure that we were not sensitive to any of the diffuse emission. Uniform weighting was used to provide the highest resolution and further suppress any diffuse emission. Based on these maps, known point sources were then carefully masked using masks generated from \textsc{PyBDSF} on the pre-point source-subtraction images. 14 faint sources were manually added to the mask at 3 GHz, 18 at 340 MHz. Figure \ref{fig:radio_pointsources} illustrates the number of sources that required masking for subtraction. 

Using the new masks, both visibilities were then deeply cleaned to create a model of only the compact emission to be subtracted. This model was then subtracted from the visibility data using \textsc{TaQL}, a command within \texttt{WSClean}. The process was repeated twice for the 340 MHz observations, and three times for the 3 GHz observations to ensure that as much compact emission as possible was subtracted out. The resulting visibilities and maps contained primarily extended emission.

\clearpage

\end{document}